\documentclass[letterpaper]{JHEP3}

\usepackage{epsfig,amsmath,euscript,graphics}

\def\Slash#1{{#1\!\!\!\slash}}

\def\bnslash{\bar n\!\!\!\slash}

\def\OMIT#1{}

\newcommand{\nn}{\nonumber}

\newcommand{\bn}{{\bar n}}
\newcommand{\bea}{\begin{eqnarray}}
\newcommand{\eea}{\end{eqnarray}}

\newcommand{\mcdot}{\!\cdot\!}

\newcommand{\muj}{\mu_j}

\newcommand{\sss}{\hat{s}}
\newcommand{\ttt}{\hat{t}}
\newcommand{\uuu}{\hat{u}}

\newcommand{\mtsq}{m^2_t}

\newcommand{\cO}{{\cal O}}

\newcommand{\la}{\langle}
\newcommand{\ra}{\rangle}

\title{Factorization and resummation of s-channel single top
quark production}

\author{Hua Xing Zhu\\
Department of Physics and State Key
Laboratory of Nuclear Physics and Technology, Peking
University, Beijing, 100871, China}

\author{Chong Sheng Li\\
Department of Physics and State Key
Laboratory of Nuclear Physics and Technology, Peking
University, Beijing, 100871, China\\
Center for High Energy Physics, Peking
University, Beijing, 100871, China\\
E-mail: \email{csli@pku.edu.cn}}

\author{Jian Wang\\
Department of Physics and State Key
Laboratory of Nuclear Physics and Technology, Peking
University, Beijing, 100871, China}

\author{Jia Jun Zhang\\
Department of Physics and State Key
Laboratory of Nuclear Physics and Technology, Peking
University, Beijing, 100871, China}

\abstract{
We study the factorization and resummation of s-channel
single top
quark production in the Standard Model at both the Tevatron
and the
LHC. We show that the production cross section in the
threshold
limit can be factorized into a convolution of hard function,
soft
function and jet function via
soft-collinear effective theory~(SCET), and resummation can
be performed using renormalization group equation in the
momentum space resummation formalism. We find that the
resummation effects significantly reduce the factorization
scale dependence of the total cross section
at the Tevatron, while at the LHC we find that the
factorization scale dependence has not been improved,
compared with the NLO results.}

\begin{document}

\section{INTRODUCTION}
\label{sec:1}

Recently, the analysis of data by D0~\cite{Abazov:2009ii}
and
CDF~\cite{Aaltonen:2009jj} collaboration has confirmed the
observation of single top production at the Tevatron. Due to
the
fact that its expected production cross section is small (
in
combined $s$ and $t$ channel, $\sigma_{st}\sim 2.9$
pb~\cite{Harris:2002md}) and the processes lack significant
signal
feature, such discovery signifies a success in both
experiment and
theory, and provides good opportunities for testing the
Standard
Model (SM) and searching for new physics.

As we know, at hadron collider, single top quarks are
produced via
three different modes: s-channel ($q\bar{q}'\to \bar{b}t$),
t-channel ($bq\to t q'$ and $b\bar{q}\to t\bar{q}'$), and
associated
$tW$ production ($bg\to tW^-$), each sensitive to quite
different
physics~\cite{Tait:2000sh}.
First of all, since all three production modes are directly
proportional to the CKM matrix element $|V_{tb}|^2$, a
measurement
of the cross section provides a unique direct probe to
$V_{tb}$, and
can constrain models with fourth generation quark. The
s-channel
single top production is rather sensitive to the interaction
mediated by extra heavy particle and the uncertainty of
partonic
luminosity for this mode is relatively small. Thus although
this
mode has the least cross section, it's a very important
channel in
searching for new physics. The t-channel is sensitive to
physics
which modifies top decay properties, and the associated
production
channel can be a good measurement of the $W-t-b$ vertex.
Thus a
precise understanding of production cross sections and
theoretical
uncertainty are important. In particular, higher order QCD
corrections to the cross section are necessary to improve
theoretical predictions. There have been a lot of NLO
calculations
of the single top production
in the
literatures~\cite{Bordes:1994ki,Stelzer:1995mi,Smith:1996ij,
Zhu:2001hw,Harris:2002md,Sullivan:2004ie,
Campbell:2004ch,Cao:2004ky,Cao:2004ap,Cao:2005pq,Cao:2008af,
Campbell:2009gj,Heim:2009ku}, with or without subsequent top
quark decay.
Recently, implementations of these results into NLO Shower
Monte
Carlo~(MC@NLO or POWHEG) also
appeared~\cite{Frixione:2005vw,Frixione:2008yi,
Alioli:2009je}.

Furthermore, the consideration of phase space logarithms in
higher order QCD effects,
which origin from incomplete
cancelation of real soft gluon emission and virtual
corrections, are also important to the theoretical
predictions.
They often dominate the hadronic cross sections and can be
systematically
resummed to all orders in perturbation
theory~\cite{Sterman:1986aj,Catani:1989ne}. For the single
top
production they have been calculated to NNLO at
Next-to-next-to-Leading-Logarithmic~(NNLL)
accuracy~\cite{Kidonakis:2010tc,Kidonakis:2010ux}, but with
some important NNLL logarithms omitted.

In recent years it has become popular to use SCET to resum
the large
phase space logarithms. SCET is an effective field theory
describing
the QCD interaction between collinear and soft
particles~\cite{Bauer:2000ew,Bauer:2000yr,Bauer:2001ct,
Bauer:2001yt}, which can correctly reproduce the long
distance
behavior of QCD, while the short distance information is
encoded in
the Wilson coefficients from matching the full theory to
SCET. In
the past decade, SCET has proved its usefulness in
describing high
energy hard scattering processes. These
 include deep-inelastic
scattering~\cite{Manohar:2003vb,Chay:2005rz,
Becher:2006nr,Chen:2006vd}, Drell-Yan
production~\cite{Idilbi:2005ky,Idilbi:2005er,Becher:2007ty,
Stewart:2009yx}, Higgs
production~\cite{Gao:2005iu,Idilbi:2005er,
Ahrens:2008qu,Ahrens:2008nc,Zhu:2009sg,Mantry:2009qz},
$e^+e^-$
annihilation  to
hadrons~\cite{Lee:2006nr,Fleming:2007qr,Fleming:2007xt,
Bauer:2008dt,Schwartz:2007ib}, color-octet scalar
production~\cite{Idilbi:2009cc}, direct photon
production~\cite{Becher:2009th}, direct top quark production
via
FCNC coupling and top quark pair
production~\cite{Yang:2006gs,Beneke:2009rj,Ahrens:2009uz,
Ahrens:2010zv}.

In this paper, we will further study the threshold
resummation
effects on the s-channel production cross section to all
orders in
QCD at NNLL accuracy in momentum space~\cite{Becher:2006nr},
utilizing SCET. {First of all, we show that the total
cross section for s-channel single top production can be
factorized
schematically as
\begin{equation}
 \sigma = f\otimes f \otimes H \otimes S \otimes J,
\end{equation}
where $f$ is the initial state nonperturbative parton
distribution
functions~(PDFs); $H$ is the hard function, which encodes
the short
distance interaction information; $S$ is the soft function,
which
describes the soft correlation between different color
objects; and
$J$ is the jet function, describing the final state
collinear
emission associated with the jet. The factorized cross
section we
derived is valid in the hadronic threshold. In this limit,
the
partons initiated the hard scattering carry almost all of
the hadron
momentum, and the final state configuration consists of a
top quark,
a narrow hard jet and the remaining soft radiations. In this
work,
we are only intereseted in the inclusive total cross
section, thus
do not consider the top quark decay effects. For this
reason, our
results can not apply to the isolated s-channel single top
cross
section measured at the Tevatron, where explicit
experimental cuts
on the final state leptons and jets are required, but only
serve as
part of the total single top production cross section. When
combined
with the results of t-channel and $tW$ associated production
channel, our results can provide the most accurate
perturbative
predictions for total cross section of single top
production. }

This paper is organized as follows. In
Section~\ref{sec:2}, we briefly review the basic
ingredients of SCET. Section \ref{sec:3} discusses the
kinematics at threshold. In section \ref{sec:4} we derive a
factorization formula for the resummed cross section in
momentum space. We give a NNLO expansion of our resummed
cross section in section \ref{sec:5}. Section \ref{sec:6}
contains a brief numerical discussion and we conclude at
section \ref{sec:7}.

\section{Brief introduction to SCET}
\label{sec:2} To describes collinear field in SCET, it is
convenient
to define a lightlike vector
$n_\mu=(1,\mathbf{n}),\mathbf{n}^2=1$.
Any four-vector can be decomposed with respect to $n_\mu$
and
$\bar{n}_\mu=(1,-\mathbf{n})$ as
\begin{equation}
 l^\mu = l^- \frac{n^\mu}{2} + l^+ \frac{\bar{n}^\mu}{2} +
l^\mu_{n\perp},
\end{equation}
with $l^+=n\mcdot l$ and $l^-=\bar{n}\mcdot l$. The momentum
of a collinear particle moving along the $n^\mu$ direction
has the following scaling
\begin{equation}
 p^\mu = (p^+, p^-, p_{n\perp} )\sim (\lambda^2, 1,
\lambda),
\end{equation}
while for a soft particle, the momentum scales as
\begin{equation}
 q\sim (\lambda^2, \lambda^2, \lambda^2),
\end{equation}
where $\lambda\ll 1$ is a small expansion parameter in SCET.
E.g.,
for an energetic jet with invariant mass $m_J$ and energy
$E_J$,
$\lambda=m_J/E_J$. From the momentum scaling, one can see
that the
interaction between collinear fields of different directions
$n_i$
and $n_j$ with $n_i\cdot n_j \gg \lambda^2$ will inevitably
change
the momentum scaling, thus is forbidden in SCET, but can
be
included as an external current in our computation. The soft
fields,
on the other hand, can interact with any collinear field
without
changing the scaling.

A $n$-collinear quark and gluon field can be written as
\begin{eqnarray}
\label{colfield}
 \chi_n(x) &=& W^\dagger_n (x) \xi_n(x),
\nn
\\
\mathcal{B}^\mu_{n\perp}(x) &=& \left[ W^\dagger_n
iD^\mu_{n\perp}W_n(x) \right],
\end{eqnarray}
where
\begin{equation}
 iD^\mu_{n\perp} = \mathcal{P}^\mu_{n\perp}+g_s
A^\mu_{n\perp}
\end{equation}
is the collinear covariant derivative and the label
operator $\cal P$ is defined to project out the large
momentum component of the collinear field, e.g., ${\cal
P}^\mu_n \xi_n = \bar{p}^\mu \xi_n$. Here we have split
$p$ into a sum of large label momentum and small residue
momentum,
\begin{equation}
 p^\mu = \bar{p}^\mu + k^\mu, \qquad \text{with}\qquad
\bar{p}^\mu = p^-\frac{n^\mu}{2} + p^\mu_{n\perp}.
\end{equation}
The collinear Wilson line,
\begin{equation}
 W_n(x) = \mathbf{P} \exp \left( ig_s \int^0_{-\infty} ds\,
\bn \mcdot A^a_n(x+s\bn)t^a \right),
\end{equation}
which describes the emission of arbitrary $n$-collinear
gluons from
an $n$-collinear quark or gluon, is constructed to make the
collinear fields as defined in Eq.~(\ref{colfield})
invariant under
the collinear gauge transformation. The operator
$\mathbf{P}$ is the
path-ordered operator acting on the color generator $t^a$.

At the leading order in $\lambda$, only the $n\mcdot A_s$
component of soft gluons can interact with the
$n$-collinear field. Such
interaction is Eikonal and can be removed by a field
redefinition~\cite{Bauer:2001yt}:
\begin{eqnarray}
\label{frd}
 \chi_n(x) &=& Y_n(x)\chi^{(0)}_n(x),
\nn
\\
 \mathcal{B}^\mu_{n\perp}(x) &=& Y_n(x)
\mathcal{B}^{\mu (0)}_{n\perp}(x) Y^\dagger_n(x),
\end{eqnarray}
where
\begin{equation}
 Y_n(x) = \mathbf{P} \exp\left( ig_s\int^0_{-\infty}ds\,
n\mcdot A^a_s(x+sn)t^a\right)
\end{equation}
for an incoming Wilson line~\cite{Chay:2004zn}. And for an
out going
Wilson line, it is defined as
\begin{equation}
 \tilde{Y}_n(x) = \mathbf{P} \exp\left(
-ig_s\int^\infty_{0}ds\, n\mcdot A^a_s(x+sn)t^a\right).
\end{equation}
The fields with superscript (0) now are decoupled with soft
gluons.
Without confusion, we will neglect the superscript below.
After the
field redefinition, the leading order SCET lagrangian is
factorized
into a sum of different collinear sectors and soft sector,
which do
not interact with each other,
\begin{equation}
\label{fl}
 \mathcal{L}_{\rm SCET} = \sum_{n_i} \mathcal{L}^{(0)}_{n_i}
+ \mathcal{L}_s + \cdots.
\end{equation}

\section{Analysis of kinematics}
\label{sec:3}

In this section, we introduce the relevant kinematical
variables
needed in our analysis. As an example, we consider the
subprocess $u
+ \bar{d} \to t + X$. The subprocesses induced by gluon
splitting
are power suppressed in the threshold limit, therefore will
not be
considered in this paper. First of all, we define two
lightlike
vectors along the beam directions, $n_a$ and $n_b$, which
are
related by $n_a=\bar{n}_b$. Then we introduce initial
collinear
fields along $n_a$ and $n_b$ to describe the collinear
particles in
the beam directions. In the center-of-mass frame of the
hadronic
collision, the momentum of the incoming hadrons can be
written as
\begin{equation}
 P^\mu_a=E_{\rm CM}\frac{n^\mu_a}{2},\qquad P^\mu_b=E_{\rm
CM}\frac{n^\mu_b}{2}.
\end{equation}
Here $E_{\rm CM}$ is the center-of-mass energy of the
collider and we have neglected the masses of the hadrons.
The
momentum of the incoming partons, with a light-cone
momentum fraction of the hadronic momentum, are
\begin{equation}
 \tilde{p}_a = x_a E_{\rm CM}\frac{n^\mu_a}{2},\qquad
\tilde{p}_b= x_b E_{\rm
CM}\frac{n^\mu_b}{2}.
\end{equation}
{
At the hadronic and partonic level, momentum conservation
means
\begin{equation}
 P_a + P_b = q + P_X,
\end{equation}
and
\begin{equation}
 \tilde{p}_a+\tilde{p}_b=q + p_X,
\end{equation}
respectively, where $q$ is the momenta of the top quark. We
define
the partonic jet with jet momentum $p_X$ to be the set of
all final
state partons except the top quark in the partonic
processes, while
the hadronic jet with jet momentum $P_X$ contains all the
hadrons as
well as the beam remnants in the final state, except the top
quark.
Explicitly, $p_X=p_1 + k$, where $p_1$ is the momentum of
the final
state collinear partons forming the jet and $k$ is the
momentum of
the soft radiations. Such division of momentum is artificial
and we
have to integrate over the soft momentum to obtain a
physical
observable. We also define the Mandelstam variables as
\begin{equation}
 s=(P_a+P_b)^2,\quad u=(P_a-q)^2, \quad t=(P_b-q)^2
\end{equation}
for hadrons, and
\begin{equation}
  \hat{s}=(\tilde{p}_a+\tilde{p}_b)^2,\quad
\hat{u}=(\tilde{p}_a-q)^2, \quad \hat{t}=(\tilde{p}_b-q)^2
\end{equation}
for partons, respectively. In terms of the Mandelstam
variables, the
hadronic and partonic threshold variables are defined as
$S_4 \equiv
P^2_X = s+t+u-m^2_t$ and $s_4 \equiv p^2_X =
\hat{s}+\hat{t}+\hat{u}
- m^2_t$, where $m_t$ is the mass of top quark. The hadronic
threshold limit is defined as $S_4 \to
0$~\cite{Laenen:1998qw}. In
this limit, the final state radiations and beam remnants are
highly
suppressed, leads to a configuration consists of a top quark
and a
narrow jet, as well as the remaining soft radiations. Taking
this
limit requires $x_a\to 1,\;x_b\to 1,\; s_4 \to 0$
simultaneously.
Thus, the hadronic threshold enforces the partonic
threshold.
However, the reverse is not true. The partonic threshold
$s_4 \to 0$
does not forbid a significant amount of beam remnants. We
note that
in both hadronic threshold limit and partonic threshold
limit, the
top quark is not forced to be produced at rest, $i.e.$ it
can have a
large momentum. For later convenience, we can also write the
threshold variable as
\begin{equation}
\label{s4a}
 s_4=p^2_X=(\tilde{p}_a+\tilde{p}_b-q)^2=p^2_1+2k^+
E_1+\cO(k^2),
\end{equation}
where $k^+=n_1\mcdot k$, $E_1$ is the energy of the quark
jet and
$n_1$ is the lightlike vector associated with the jet
direction.
Note that our definition of $s_4$ is different
from~\cite{Kidonakis:2010tc}, in which the definition
$\bar{s}_4=(\tilde{p}_a+\tilde{p}_b-p_1)^2 - m^2_t$ is
adopted,
where we put a bar on $s_4$ to distinguish the definition
in~\cite{Kidonakis:2010tc} from ours. We point out that
the meaning of such choice is not clear, because $\bar{s}_4$
doesn't vanish when there are collinear gluon emitting from
the final state b-quark. However, as we know, there are
large logarithms associated with such collinear
gluon emission, thus, the definition adopted
in~\cite{Kidonakis:2010tc} could miss some large
contributions.}

In the
threshold limit~($s_4\to 0$), incomplete cancelation between
real
and virtual corrections leads to singular distributions
$\alpha^n_s
[\ln^m(s_4/m^2_t)/s_4]_+$, with $m \leq 2n-1$. It is the
purpose of
threshold resummation to sum up these contributions to all
orders in
perturbation theory.

\section{Factorization in SCET}
\label{sec:4}

In this section, we derive the factorized cross section
formula for
s-channel single top production in SCET, following the
convention
and formalism of~\cite{Bauer:2008jx,Bauer:2010vu}. With
appropriate changes, our
formula can be extended to the resummation of t-channel and
$tW$
associated production channel as well, which will be
presented
elsewhere~\cite{wang,zhang}.

Total cross section for s-channel single top production can
be written
as~\cite{Bauer:2008jx}:
\begin{eqnarray}
\label{eq1}
 \sigma &=& \frac{1}{2E^2_{\rm CM}}
\sum^{\rm res.}_X \la I | O^\dagger_x(0) | X \ra \la X|
O_x(0)
| I\ra (2\pi)^4\delta^4(P_a+P_b-p_X)
\nn
\\
&=& \frac{1}{2E^2_{\rm CM}}
\sum^{\rm res.}_X \int d^4x\la I | O^\dagger_x(x) | X \ra
\la X|
O_x(0)
| I\ra
\nn
\\
&=& \frac{1}{2E^2_{\rm CM}}
\sum^{\rm res.}_X 
\int d^4x\int \frac{d^4 k}{(2\pi)^4} e^{-ik\cdot x} \int
\frac{d^4 p}{(2\pi)^4} 
\la I | O^\dagger(k) | X \ra \la X|
O(p)
| I\ra
\nn
\\
&=& \frac{1}{2E^2_{\rm CM}} \sum^{\rm res.}_X
\int\,\frac{d^4p}{(2\pi)^4}
\la I| O^\dagger(0) | X \ra \la X| O(p)| I
\ra ,
\end{eqnarray}
where $|I\ra=|P_a P_b\ra$ denotes the initial state protons
(anti-protons), $O_x(x)$ is the relevant operator responsible for
the underlying hard interaction, and $O(p)$ its fourier
transformation. Here we distinguish the position space operator from
the momentum space one by a subscript $x$. The restriction on the
sum over final states $|X\ra$ is that we include final state
configuration consists only of a top quark jet whose 3-momentum is
in the direction of $\bar{n}_1$, an anti b quark quark jet in the
direction of $n_1$, and soft radiations. This is the configuration
that is relevant to threshold resummation and we are interested in.
Under this condition the final state can be written as $|X\ra = |X_t
X_1 X_s\ra$, where $|X_t\ra$, $|X_1\ra$ and $|X_s\ra$ denote top
quark jet, anti b quark jet and remaining soft radiations,
respectively. In the second line of Eq.~(\ref{eq1}), we have used
the momentum conservation delta function to shift the operator
$O_x^\dagger$ to point $x$, and in the third line written
the operators in momentum
space, which are then matched onto SCET operators. We chose to match
directly in momentum space, which has the advantage that the
cumbersome label summations and residual soft integration can be
combined to a full four momentum integral~\cite{Bauer:2008jx}. After
matching we obtain
\begin{eqnarray}
\label{ope}
 O(p) &=& \int\,
\frac{d^4p_a}{(2\pi)^4}
\frac{d^4p_b}{(2\pi)^4} \frac{d^4p_1}{(2\pi)^4}
\frac{d^4p_2}{(2\pi)^4}
\frac{d^4k_s}{(2\pi)^4}\mathcal{C}_I(p_a,p_b;p_1,p_2)
\nn
\\
&&
\times \cO_{in}(p_a,p_b) \cO_{out} (p_1,p_2)
\cO_{S,I}(k_s)(2\pi)^4\delta^{(4)}(p-p_a-p_b+p_1+p_2+k_s).
\end{eqnarray}
In the above equation, we have made all the spin,
Lorentz and color indices implicit. For example, at the LO,
the hard matching coefficient $\mathcal{C}_I$ reads
\begin{equation}
\label{lohmc}
\mathcal{C}_I=
i\frac{\alpha\pi V^*_{ud}V_{tb}}{2\sin^2\theta_W
((p_a+p_b)^2-M^2_W)}(\gamma^\mu(1-\gamma^5))^{\gamma\delta}
(\gamma_\mu(1-\gamma^5))^{\alpha\beta} \delta_{I1},
\end{equation}
where $\alpha$ is the fine-structure constant, $V_{ij}$ the
CKM
matrix element, $\theta_W$ the electroweak mixing angle, and
$M_W$
the mass of the $W$ boson. Here we have chosen the s-channel
singlet-octet basis as the independent color structure for
this
process:
\begin{equation}
 |c_1\ra = \delta_{cd}\delta_{ef},\qquad
|c_2\ra=(t^a)_{cd}(t^a)_{ef},
\end{equation}
and $I=1$ or $2$ is an index in this color space. Thus the
kronecker
delta function in Eq.~(\ref{lohmc}) can be understood that
at the LO
the final state $t\bar{b}$ pair of s-channel single top
production
is a color-singlet. In Eq.~(\ref{ope}) $\mathcal{O}_{in}$
denotes
the effective operator responsible for annihilating an
initial state
collinear up-type quark with momentum $p_a$ and an down-type
anti-quark with momentum $p_b$, which can be written as
\begin{equation}
\label{isf}
 \cO^{cd}_{\alpha\beta,in} = \bar{\chi}^c_\alpha(-p_b)
\chi^d_\beta(p_a),
\end{equation}
 and $\mathcal{O}_{out}$ is the effective operator
responsible for the creation of final state anti-bottom
quark with
momentum $p_1$ and top quark with momentum $p_2$, which can
be
expressed as
\begin{equation}
\label{fsf}
 \cO^{ef}_{\gamma\delta,out} =
\bar{h}^e_{\gamma,v}(p_2) \chi^f_\delta (-p_1).
\end{equation}
Note that we have taken the $m_t \to \infty$ limit at fixed
top jet radius and then described the top quark in terms of
the
heavy
quark effective field~\cite{Isgur:1989vq} with label
velocity $v$.
The fields in Eqs.~(\ref{isf}) and (\ref{fsf}) are defined
with
field redefinition in Eq.~(\ref{frd}), thus they no
longer
interact with soft degrees of freedom.

The soft operators $\cO_{S,I}$, which are responsible for
the soft
interactions between different collinear sectors and top
quark, are
expressed as
\begin{eqnarray}
 \cO_{S,1}(k_s)&=&\int\,d^4x
e^{-ik_s\,\mcdot \,x} \mathbf{T}\left[ \left(
Y^\dagger_{n_b} (x) Y_{n_a}(x) \right)^{cd}
\left(\tilde{Y}^\dagger_{v_2}(x)\tilde{Y}_{n_1}
(x)\right)^{ef}\right],
\nn
\\
 \cO_{S,2}(k_s)&=&\int\,d^4x
e^{-ik_s\,\mcdot \,x} \mathbf{T}\left[ \left(
Y^\dagger_{n_b} (x)
t^a Y_{n_a}(x) \right)^{cd}
\left(\tilde{Y}^\dagger_{v_2}(x)t^a\tilde{Y}_{n_1}
(x)\right)^{ef}\right],
\end{eqnarray}
where the time-ordering operator $\mathbf{T}$ is required to
ensure
the proper ordering of soft gluon fields in the soft Wilson
line.

Using the notation $\Phi_2=\{p_a,p_b;p_1,p_2\}$ to express a
phase
space point~\cite{Bauer:2008jx} with $d\Phi_2=d^4 p_a d^4
p_b d^4p_1 d^4p_2/(2\pi)^{16}$ and $\Phi_2 - k_s = p_a +
p_b - p_1 - p_2 - k_s$,  we can write
Eq.~(\ref{eq1}) in a
compact form
\begin{eqnarray}
\label{fdcs}
  \sigma&=&
\frac{1}{2E^2_{\rm CM}} \sum^{res}_X
\int d\Phi'_2 d\Phi_2 \mathcal{C}^*_J(\Phi'_2)
\mathcal{C}_I(\Phi_2)\int
\frac{d^4k'_s}{(2\pi)^4}\,\frac{d^4k_s}{(2\pi)^4}
(2\pi)^4\delta^{(4)}(\Phi_2-k_s)
\nn
\\
&&
\times
\la I | (\cO'_{in}\cO'_{out}\cO'_{S,J})^\dagger |X_t
X_1 X_s\ra\la X_t X_1 X_s|
(\cO_{in}\cO_{out}\cO_{S,I}) |I\ra.
\end{eqnarray}
As we mentioned before, different collinear sectors are
decoupled
due to field redefinition, and thus the matrix element in
Eq.~(\ref{fdcs}) can be factorized into a product of several
matrix
elements, which obey certain renormalization group (RG)
equation.

In the following, we further show the matrix elements
mentioned
above. First, we deal with the top quark sector. Since we
have
decoupled the soft interaction by field redefinition, the
top quark
now should be regarded as a non-interacting particle, which
can be
written as
\begin{eqnarray}
 &&\sum_{X_t}\int \frac{d^4 p'_2}{(2\pi)^4}\frac{d^4
p_2}{(2\pi)^4}
\la 0| {h^{e'}_{\gamma',v'_2}} (p'_2) |X_t\ra \la X_t|
\bar{h}^e_{\gamma,v_2}(p_2) | 0 \ra...
\nn
\\
=&& \int\frac{d^3 q}{2 E_q(2\pi)^3}(\Slash{q} +
m_t)_{\gamma'
\gamma}\delta_{e'e}...
\end{eqnarray}
where summation over final state $|X_t\ra$ gives rise to a
top quark phase space integral.
Next, we define the soft function by the soft matrix element
as
\begin{eqnarray}
 S^{f'e'd'c'cdef}_{JI}(k^+,\mu) &=& \sum_{X_s}
 \int dk^+ \frac{d^4 k'_s}{(2\pi)^4}\frac{d^4
k_s}{(2\pi)^4} \la 0|
{\mathcal{O}^{\dagger,f'e'd'c'}_{S,J}}(k'_s)
\nn
\\
&&
\delta[k^+ -n_1 \mcdot k_s] |X_s
\ra\la X_s |\mathcal{O}^{cdef}_{S,I} (k_s)|
0 \ra,
\end{eqnarray}
where we have inserted into the above equation an identity
operator
\begin{equation}
 \mathbf{1}=\int dk^+ \, \delta[k^+ - n_1 \mcdot
k_s].
\end{equation}
Note that the summation over final state can be performed
$\sum_{X_s} |X_s\ra \la X_s|=1$ since there is no
restriction in the summation when written between the soft
operator and also there is no explicit dependence on
$|X_s\ra$.
Since we are only interested in the total cross sections,
the final state top quark, jet function and
PDFs can be considered to be diagonal in color space. Then
we can contract them to obtain
the soft function matrix
\begin{equation}
  S_{JI}(k^+,\mu) = \delta^{f'f}
\delta^{e'e} \delta^{d'd} \delta^{c'c}
S^{f'e'd'c'cdef}_{JI}(k^+,\mu).
\end{equation}
At the LO, it can be written as
\begin{equation}
 \mathbf{S}(k^+,\mu)=\delta(k^+)
\left(
\begin{array}{cc}
N^2_c & 0\\
0 & \frac{N^2_c -1}{4}
\end{array}
\right).
\end{equation}
{At the NLO, the calculation of soft function
boils down to
the evaluation of eikonal diagrams~\cite{Becher:2009th}.
Since the virtual corrections in SCET vanish, only real
emission diagrams, which are shown in Figs.~\ref{softint},
are needed to be evaluated. The detail calculation of these
diagrams are given in the appendix.}

\FIGURE{
\centering
\begin{tabular}{ccc}
\epsfig{file=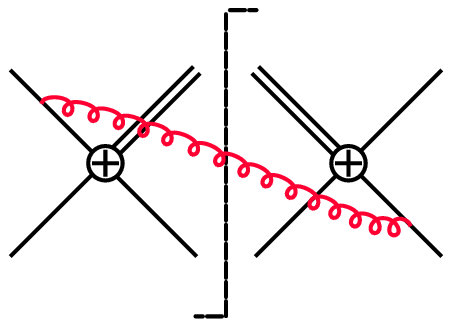,width=0.25\linewidth,clip=}
&
\epsfig{file=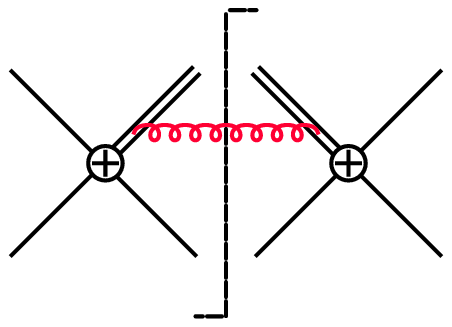,width=0.25\linewidth,
clip= }
&
\epsfig{file=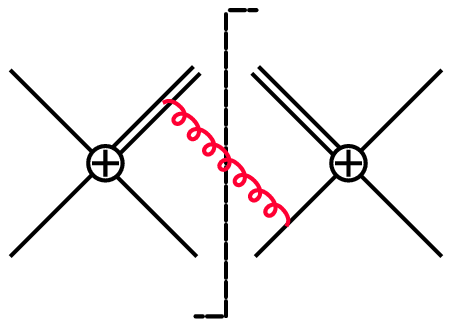,width=0.25\linewidth,
clip= }
\\
(a)&(b)&(c)
\\
\epsfig{file=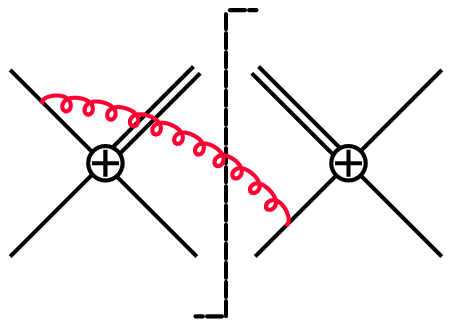,width=0.25\linewidth,clip=}
&
\epsfig{file=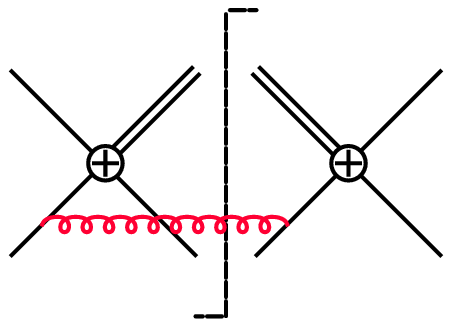,width=0.25\linewidth,clip=}
&
\epsfig{file=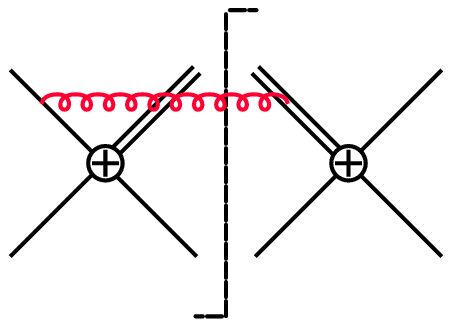,width=0.25\linewidth,clip=}
\\
(d)&(e)&(f)
\\
&
\epsfig{file=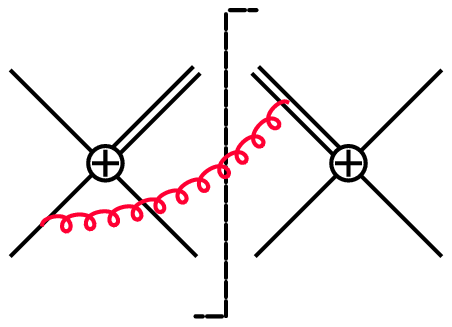,width=0.25\linewidth,clip=}
&
\\
&(g)&
\end{tabular}
\caption{Diagrams contribute to the soft function. the
double line represents the top quark line. Diagrams
(a), (c), (d), (e), (f) and (g) have their mirror image
diagrams. There are also diagrams with soft gluon
connecting to the same massless quark line on the two side
of the cut, which are not shown here because they vanish
identically. The cross dot in the diagrams represents either
singlet or octet operator.}
\label{softint}
}

For the final state anti-quark jet sector, we
have~\cite{Bauer:2008jx}
\begin{eqnarray}
\label{jet1}
&& \sum_{X_1}\int
\frac{d^4p'_1}{(2\pi)^4}\frac{d^4p_1}{(2\pi)^4}
\la 0| \bar{\chi}^{f'}_{\delta'}(-p'_1) |X_1\ra \la X_1|
\chi^f_\delta (-p_1) |0\ra
\nn
\\
&&=
\delta^{f'f}
\int\frac{d^4p_1}{
(2\pi)^3 }
\left(\frac{\Slash{n}_1}{2}\right)_{\delta\delta'}
\theta(p^0_1) p^-_1
J(p^2_1),
\end{eqnarray}
Again, summation over collinear state has been performed
and $J$ is the spin and color singlet jet function,
which
can be
defined as
\begin{equation}
\theta(p^0) p^- J(p^2)=\frac{1}{8\pi N_c}
\int \frac{d^4p'}{(2\pi)^4} {\rm Tr}
\la 0|  \bar{\chi}(-p') \bnslash_1
\chi (-p) |0\ra.
\end{equation}
Finally, the initial state collinear sector reduces to the
conventional PDFs~\cite{Bauer:2001yt}, of which the matrix
element
for $n_a$ direction is
\begin{eqnarray}
 &&\int  \frac{d^4p'_a}{(2\pi)^4}
\frac{d^4p_a}{(2\pi)^4} \la P_a | \bar{\chi}^{d'}_{\beta'}
(p'_a)  \chi^d_\beta
(p_a) | P_a \ra
=
\frac{1}{2 N_c}\delta^{d'd} \int^1_0 \frac{dx_a}{x_a}
\left(x_a E_{\rm CM}\frac{\Slash{n}_a}{2}\right)_{\beta
\beta'} f( x_a, \mu),
\end{eqnarray}
and similarly for the matrix element for $n_b$ direction.
Thus
the momentum of incoming partons are given by
$\tilde{p}_{a,b}=x_{a,b}E_{\rm CM}n^\mu_{a,b}/2$.

Combining the above expressions, we obtain (up to power
corrections)
\begin{eqnarray}
\label{fe}
 \sigma &=&
\frac{1}{2E^2_{\rm CM}} \frac{1}{4N^2_c}\int^1_0
\frac{dx_a}{x_a}\frac{dx_b}{x_b} \int  \frac{d^3q}{2 E_q
(2\pi)^3}
 f_{i/P_a}(x_a,\mu_f)
f_{j/P_b}(x_b,\mu_f)
 \lambda_{0,ij} H_{IJ}
\nn
\\
&& \times
\int dk^+\, S_{JI}(k^+_i,\mu)(2\pi) J(s_4-2k^+
E_1,\mu),
\end{eqnarray}
with
\begin{equation}
 \lambda_{0,ij} =  \frac{e^4}{\sin^4\theta_W}
|V_{ij}|^2
|V_{tb}|^2 \frac{(\hat{t}-m^2_t)\hat{t}}{(\hat{s}-M^2_W)^2}.
\end{equation}
At the LO, the hard function $H_{IJ}$ is normalized to
$\delta_{I1}\delta_{J1}$. In general, it is related to the
amplitudes of full theory by~\cite{Ahrens:2010zv}
\begin{eqnarray}
 \lambda_{0,ij} H^{(0)}_{IJ} &=& \frac{1}{\la c_I|c_I\ra \la
c_J
| c_J \ra} \la c_I | \mathcal{M}^{(0)}_{\rm ren} \ra \la
\mathcal{M}^{(0)}_{\rm ren} | c_J \ra,
\nn
\\
\lambda_{0,ij} H^{(1)}_{IJ} &=& \frac{1}{\la c_I|c_I\ra \la
c_J |
c_J \ra} \left(\la c_I | \mathcal{M}^{(1)}_{\rm ren} \ra
\la
\mathcal{M}^{(0)}_{\rm ren} | c_J \ra+\la c_I |
\mathcal{M}^{(0)}_{\rm ren} \ra \la
\mathcal{M}^{(1)}_{\rm ren} | c_J \ra\right),
\end{eqnarray}
where $|\mathcal{M}_{\rm ren} \ra$ are obtained by
subtracting the
IR divergences in the $\overline{\rm MS}$ scheme from the UV
renormalized amplitudes of full theory. {To 1-loop
order, it reduces to evaluating in the full theory the
1-loop
Feynman diagrams in Fig.~\ref{box}. The complete 1-loop hard
function
is shown in the appendix.}
\FIGURE{
\centering
 \includegraphics[]{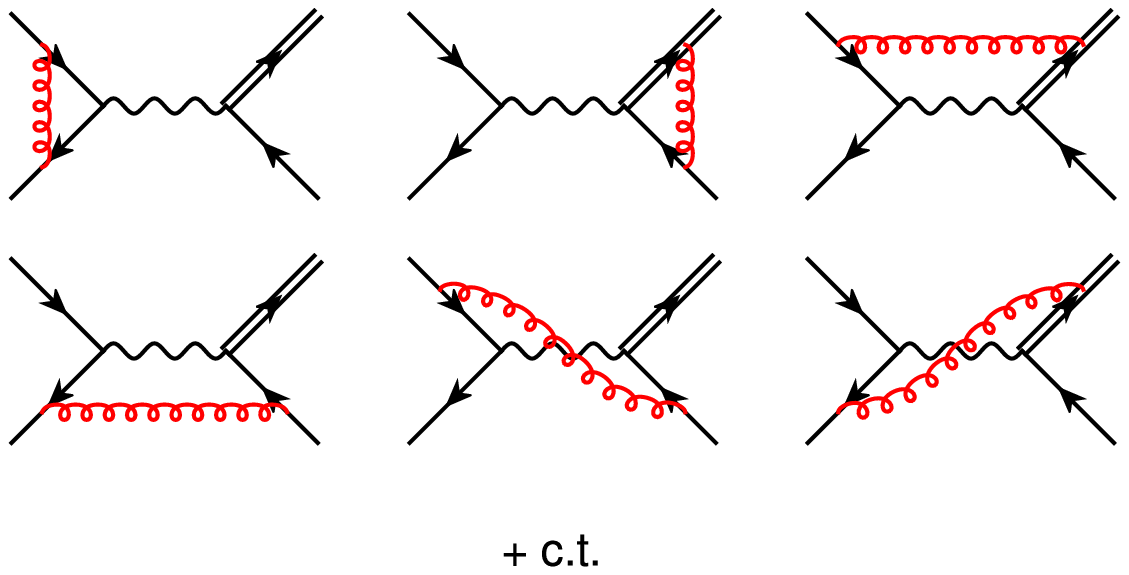}
\caption{1-loop Feynman diagrams for the hard function. The
double line represents the top quark.}
\label{box}
}

The hard function is a $2\times 2$ matrix in the color
space. The RG
equation it obeys reads
\begin{equation}
\label{hardrg}
 \frac{d\, \mathbf{H}}{d\ln \mu} = \mathbf{\Gamma}_H
\mathbf{H} +  \mathbf{H} \mathbf{\Gamma}^\dagger_H.
\end{equation}
The relevant anomalous dimension matrix is process
dependent and can be expanded in $\frac{\alpha_s}{4\pi}$,
with the relevant expansion coefficient given in the
appendix. The explicit form of the anomalous dimension
matrix for s-channel single top production can be extracted
from a more
general result given in the Ref.~\cite{Becher:2009kw}.
Explicitly,
for the independent color basis we have chosen, it is given
by
\begin{equation}
 \mathbf{\Gamma}_H=\left[\frac{3}{2} C_F \gamma_{\rm
cusp}(\alpha_s)
\ln\frac{m^2_t}{\mu^2} +\bar{\gamma}_h(\alpha_s)
\right]\mathbf{1}+\mathbf{\Gamma}_h,
\end{equation}
with
\begin{equation}
 \bar{\gamma}_h(\alpha_s) = C_F\gamma_{\rm cusp}(\alpha_s)
\ln\frac{\hat{s}(\hat{s}-m^2_t)}{m^4_t} +
\gamma_h(\alpha_s),
\end{equation}
and
\begin{equation}
 \mathbf{\Gamma}_h=\gamma_{\rm cusp}(\alpha_s)
\left(
\begin{array}{cc}
0 & \frac{C_F}{2N_c}
\ln\frac{\hat{u}(\hat{u}-m^2_t)}{\hat{t}(\hat{t}-m^2_t)}\\
\ln\frac{\hat{u}(\hat{u}-m^2_t)}{\hat{t}(\hat{t}-m^2_t)} &
\frac{N_c}{2}\ln\frac{\hat{u}
(\hat { u }
-m^2_t)}{\hat{s}(\hat{s}-m^2_t)}-\frac{1}{N_c}\ln\frac{\hat{
u }
(\hat { u } -m^2_t) } { \hat{t}(\hat{t}-m^2_t)}
\end{array}
\right).
\end{equation}
The expression for $\gamma_h$ and $\gamma_{\rm cusp}$ can
be found in the appendix.
Note that we have only retained the real part of the
anomalous
dimension matrix. {We have checked that the 1-loop hard
function exactly obeys this equation, as it must. Details of
it are
presented in the appendix.}

The solution of the RG equation shown in Eq.~(\ref{hardrg})
can be
obtained by diagonalizing the anomalous dimension matrix
$\mathbf{\Gamma}_h$~\cite{PhysRevD.64.114001}. A detail
example for how to do this in effective field theory can be
found in the Ref.~\cite{Ahrens:2010zv}. To NNLL accuracy,
the result can be expressed as
\begin{eqnarray}
 \mathbf{H}_R (\mu) &=& \exp[
6S_{\rm DL}(\mu_{h},\mu)-2\bar{a}_{h}(\mu_{h},\mu)]
\left(\frac{m^3_t}{\mu^3_h} \right)^ {
-2a_\Gamma
(\mu_{h},\mu)}
\nn
\\
&&
\times
\mathbf{U}_R(\mu_h,\mu)
\mathbf{H}(\mu_h)
 \mathbf{U}^\dagger_R(\mu_h,\mu),
\end{eqnarray}
with
\begin{equation}
\mathbf{U}_R(\mu_h,\mu)=(\mathbf{1}+\frac{\alpha_s(\mu)}{
4\pi} \mathbf{J} ) \left(
\frac{\alpha_s(\mu_h)}{\alpha_s(\mu)}
\right)^{\mathbf{\Gamma}^0_h/(2\beta_0)}
(\mathbf{1}-\frac{\alpha_s(\mu_h)}{
4\pi} \mathbf{J} ),
\end{equation}
where
\begin{equation}
 \mathbf{J} = \frac{\beta_1}{2\beta^2_0} \mathbf{\Gamma}^0_h
- \frac{1}{2\beta_0} \mathbf{\Gamma}^1_h.
\end{equation}
The Sudakov exponent are given by~\cite{Becher:2007ty}:
\begin{eqnarray}
 S_{\rm DL} (\nu,\mu) &=&
-\int^{\alpha_s(\mu)}_{\alpha_s(\nu)}\,d\lambda
\frac{C_F\gamma_
{ \rm
cusp} (\lambda)}{\beta(\lambda)}
\int^{\lambda}_{\alpha_s(\nu)}\,\frac{d\lambda'}{
\beta(\lambda')
},
\nn
\\
a_\Gamma (\nu,\mu) &=&
-\int^{\alpha_s(\mu)}_{\alpha_s(\nu)}\,d\lambda
\frac{C_F\gamma_
{ \rm
cusp}(\lambda)}{\beta(\lambda)},
\nn
\\
\bar{a}_h (\nu,\mu) &=&
-\int^{\alpha_s(\mu)}_{\alpha_s(\nu)}\,d\lambda\frac{
\bar{\gamma}_h(\lambda)}{\beta(\lambda)}.
\end{eqnarray}

Next, we discuss the jet function. Unlike the hard function
which
obeys a local RG equation, the jet function $J(p^2,\mu)$
satisfies a
RG equation which is non-local in
$p^2$~\cite{Becher:2006mr}:
\begin{eqnarray}
 \frac{dJ(p^2,\mu)}{d\ln\mu} &=& \left( -2 C_F\gamma_{\rm
cusp}(\alpha_s)
\ln\frac{p^2}{\mu^2} - 2 \gamma_j
(\alpha_s)\right)J(p^2,\mu)
\nn
\\
&&
+2C_F\gamma_{\rm cusp}(\alpha_s)\int^{p^2}_0
dq^2\,\frac{J(p^2,\mu)-J(q^2,\mu)}{p^2-q^2}.
\end{eqnarray}
This equation is solved with the help of Laplace
transformed jet function~\cite{Becher:2006mr}:
\begin{equation}
 \widetilde{j}(\ln\frac{Q^2}{\mu^2},\mu)=\int^\infty_0
dp^2\,\exp(-\frac{p^2}{Q^2e^{\gamma_E}}) J(p^2,\mu),
\end{equation}
which satisfies the RG equation
\begin{equation}
 \frac{d\,\widetilde{j}(\ln\frac{Q^2}{\mu^2},\mu)}{d\ln\mu}
=\left(-2
C_F\gamma_{\rm cusp}(\alpha_s)
\ln\frac{Q^2}{\mu^2}-2\gamma_j(\alpha_s)\right)\tilde{j}
(\ln\frac{Q^2} {\mu^2},\mu),
\end{equation}
and can be easily solved now. The solution, after
transformed back
to momentum space, is~\cite{Becher:2006mr}
\begin{equation}
 {J}(p^2,\mu)=\exp \left( -4S_{\rm
DL}(\mu_j,\mu)+2a_j(\mu_j,\mu)
\right) \widetilde{j}
(\partial_{\eta_j}, \mu_j )  \frac{1}{p^2} \left(
\frac{p^2}{\mu^2_j}\right)^{\eta_j}
\frac{e^{-\gamma_E
\eta_j}}{\Gamma(\eta_j)},
\end{equation}
where $\eta_j=2 a_\Gamma(\muj,\mu)$.

{Finally, we need the RG equation of the soft
function,
which can be obtained by noticing the fact that the hadronic
cross
section in the threshold region should be independent of the
arbitrary scale $\mu$. Schematically,
\begin{equation}
 \frac{d}{\ln\mu} f\otimes f \otimes \textbf{H} \otimes
\textbf{S} \otimes J =0.
\end{equation}
Based on this fact, we have}
\begin{eqnarray}
\label{softlp}
\widetilde{\mathbf{s}}\left(\ln\frac{\kappa}{\mu},
\mu\right)&=&\int^\infty_0 dk^+
\exp\left(-\frac{k^+}{\kappa e^{\gamma_E}} \right)
\mathbf{S}(k^+,\mu),
\\
\label{sevol}
 \frac{d\,\widetilde{\mathbf{s}}\left(\ln\frac{\kappa}{\mu},
\mu\right)}{d\ln\mu}
&=&-\mathbf{\Gamma}^\dagger_S
\widetilde{\mathbf{s}}\left(\ln\frac{\kappa}{\mu},
\mu\right) -
\widetilde{\mathbf{s}}\left(\ln\frac{\kappa}{\mu},
\mu\right) \mathbf{\Gamma}_S,
\end{eqnarray}
with
\begin{eqnarray}
\mathbf{\Gamma}_S &=&  \left(C_F\gamma_{\rm cusp}(\alpha_s)
\ln\frac{\kappa}{\mu} - \bar{\gamma}_s(\alpha_s)\right)\mathbf{1} +
\mathbf{\Gamma}_h,
\end{eqnarray}
where
\begin{equation}
 \bar{\gamma}_s(\alpha_s) = -C_F \gamma_{\rm cusp}(\alpha_s)
\ln\frac{2 n_a\mcdot n_b\, v\mcdot
n_1}{n_a\mcdot n_1 \,
n_b\mcdot n_1} + \gamma_s (\alpha_s),
\end{equation}
and $\gamma_s$ is given in the appendix.
Note that in obtaining the RG
equation Eq.~(\ref{sevol}), we have used the DGLAP evolution
for the
PDFs in the $x\to 1$  limit:
\begin{equation}
\label{DGLAP}
 \frac{df_{q/N}(x,\mu)}{d\ln\mu}=2\gamma_\phi (\alpha_s)
f_{q/N}(x,\mu) + 2 C_F\gamma_{\rm cusp} (\alpha_s) \int^1_x
\frac{dz}{z} \frac{f_{q/N} (x/z,\mu)}{[1-z]_+}.
\end{equation}
{We stress that Eq.~(\ref{sevol}) is derived
entirely from
RG invariance of the resummed cross section in the threshold
limit.
If the explicit 1-loop soft function do obey this equation,
it
serves as a non-trivial check on the RG invariance of our
result. We
show in the appendix that this is indeed the case, as
expected.} The
solution of the RG equation in momentum space is
\begin{eqnarray}
 \mathbf{S}_R(k^+,\mu) &=& \exp [-2S_{\rm DL}
(\mu_s,\mu) - 2\bar{a}_s(\mu_s,\mu)]
\nn
\\
&& \times \mathbf{U}^\dagger_R(\mu,\mu_s)
\widetilde{\mathbf{s}}(\partial_{\eta_s},\mu_s)
\mathbf{U}_R(\mu,\mu_s)
\frac{1}{k^+} \left( \frac{ k^+}{\mu_s} \right)^{\eta_s}
\frac{e^{-\gamma_E \eta_s}}{\Gamma(\eta_s)},
\end{eqnarray}
where $\eta_s=2a_\Gamma(\mu_s,\mu)$.
Combining the above ingredients, we obtain the resummed
cross section for s-channel single top production
\begin{eqnarray}
\label{diffxec}
 \sigma^{\rm thres} &=&
\sum_{ij}\frac{\pi}{4
N^2_cE^2_{\rm CM}}
\int^1_0 \frac{dx_a}{x_a} \frac{dx_b}{x_b} \int \frac{d^3
q}{2 E_q (2\pi)^3}
f_{i/N_a}(x_a)f_{j/N_b}(x_b) \lambda_{0,ij}
\nn
\\
&& \times\int^{s_4/(2E_1)}_0 dk^+\,\mathrm{Tr}[
\mathbf{H}_R(\mu)
\mathbf{S}_R(k^+,\mu)] J(s_4-2E_1k^+,\mu),
\end{eqnarray}
where we have included a summation over different partonic
channels.
\section{NNLO expansion of resummed cross section}
\label{sec:5}

In the traditional approach to threshold resummation, the
evolution
equations for the factorized cross section are solved in
Mellin
moment space rather than momentum space. It has been
demonstrated
that in Drell-Yan production, the two approach are
equivalent up to
$1/N$ corrections when making the scale choices $\mu_h=M$
and
$\mu_s=M/N$~\cite{Becher:2007ty}. Here $N$ is the Mellin
moment and
$M$ is the mass of the Drell-Yan lepton pair. Note that when
$N$ is
very large, the running coupling constant $\alpha_s(\mu_s)$
blows
up. Furthermore, in order to obtain the physical cross
section in
momentum space, one needs to invert the Mellin
transformation
numerically, which is ambiguous since the expression to be
inverted
has a Landau-pole for large $N$, although there have been
several
prescriptions for dealing with the Mellin
inversion~\cite{Laenen:1991af,Berger:1996ad,Catani:1996yz}.
On the
other hand, in the momentum space approach, we do not
encounter such
problem because $\mu_s$ is constrained to be well above the
Landau-pole singularity.

It has also been advocated that the resummed cross section
in Mellin
moment space can be used as a generator of the fixed order
perturbation expansion~\cite{Kidonakis:2000ui}. The
ambiguity due to
different Mellin inversion prescriptions can be avoided
because at
the fixed order no prescription is needed to invert the
Mellin space
results. In this way, partial NNNLO threshold singular terms
were
obtained for s-channel single top
production~\cite{Kidonakis:2006bu,Kidonakis:2007ej,
Kidonakis:2010tc}.

In order to obtain the fixed order expansion in momentum
space
resummation formalism, we rewrite the cross section in terms
of
integral over singular distributions of $s_4$. In the
center-of-mass
frame of the top quark and the recoiling jet, we can
parametrize the
momentum of $\tilde{p}_a$, $\tilde{p}_b$ and $q$
as~\cite{Beenakker:1988bq}
\begin{equation}
 \tilde{p}_a = \frac{\sqrt{\hat{s}}}{2}(1,0,0,1),\quad
\tilde{p}_b = \frac{\sqrt{\hat{s}}}{2}(1,0,0,-1),\quad
q=(E_q, 0, |\mathbf{q}|\sin\chi, |\mathbf{q}|\cos\chi),
\end{equation}
where
\begin{equation}
 E_q=-\frac{\hat{t}+\hat{u}-2m^2_t}{2\sqrt{\hat{s}}},\quad
\cos\chi=\frac{\hat{u}-\hat{t}}{\sqrt{(\hat{t}+\hat{u}
-2m^2_t)^2-4\hat{s}m^2_t } }.
\end{equation}
Hence the phase space measure of the top quark is
\begin{eqnarray}
 \frac{d^3 q}{2E_q (2\pi)^3}&=&\frac{1}{8\pi^2}dE_q
d\chi\,\sin\chi\sqrt{E^2_q-m^2_t}
\nn
\\
&=& \frac{1}{16\pi^2\hat{s}}d\hat{t}d\hat{u},
\end{eqnarray}
and the cross section can now be rewritten as
\begin{eqnarray}
\sigma &=& \int^1_{m^2_t/s} d\tau \int^1_\tau
\frac{dx_a}{x_a}  \int^0_{m^2_t-\hat{s}} d\hat{t}
\int^{s^{\rm max}_4}_0 ds_4\, f_{i/N_a}(x_a,\mu_F) f_{j/N_b}
(\tau/x_a,\mu_F)
\frac{d\hat{\sigma}^{\rm thres}}{d\hat{t}d\hat{u}},
\end{eqnarray}
where $s^{\rm max}_4=\hat{s}+\hat{t}+m^2_t
\hat{s}/(\hat{t}-m^2_t)$, and
\begin{eqnarray}
\frac{d\hat{\sigma}^{\rm thres}}{d\hat{t}d\hat{u}} &=&
\sum_{ij}\frac{\lambda_{0,ij}}{64\pi N^2_c\hat{s}^2}
\int^{s_4/(2E_1)}_0
dk^+\,\mathrm{Tr}[ \mathbf{H}_R(\mu)
\mathbf{S}_R(k^+,\mu)]
J(s_4-2E_1k^+,\mu).
\end{eqnarray}
Following~\cite{Becher:2007ty}, we derive the threshold
singular
distributions by setting $\mu_h$, $\mu_s$ and $\mu_j$ equal
to the
common scale $\mu$, which is conveniently chosen as the
factorization scale $\mu_F$. In the following, we show all
the threshold
singular distributions up to NNLO,
\begin{eqnarray}
\label{thresexp}
 &&\frac{64\pi\hat{s}^2 d^2\hat{\sigma}^{\rm
expand}_{ij}}{\lambda_{0,ij} d\hat{t}\,d\hat{u}} =
\delta(s_4) +
\frac{\alpha_s}{4\pi}\left( A_2 D_2 + A_1D_1
+A_0 \delta(s_4)\right)
\nn
\\
&&
+ \left( \frac{\alpha_s}{4\pi} \right)^2\left(
B_4 D_4 + B_3 D_3 + B_2 D_2 + B_1 D_1 + B_0
\delta(s_4)\right),
\label{nnlo_expanded}
\end{eqnarray}
where
\begin{equation}
 D_n = \left[\frac{1}{s_4}\ln^{n-1}\frac{
s_4}{m^2_t}\right]_+
\end{equation}
is the conventional plus distribution, and its integral with a
regular function $f(s_4)$ is defined as
\begin{equation}
 \int^{\mtsq}_0 ds_4\, f(s_4) D_n = \int^{\mtsq}_0 ds_4\,
[f(s_4)-f(0)]\frac{1}{s_4}\ln^{n-1}\frac{s_4}{\mtsq}.
\end{equation}
The coefficients $A_n$ and $B_n$ are given by
\begin{eqnarray}
 A_2 &=& 3C_F\gamma^0_{\rm cusp},
\\
A_1 &=& C_F \gamma^0_{\rm cusp} (L_h + 2L_s)+\gamma^0_j
-2\bar{\gamma}^0_s,
\\
A_0 &=& \frac{c^s_{11}}{C^2_A} + c^h_{11} + c^j_1 + C_F
\gamma^0_{\rm cusp} \left( -\frac{L^2_h}{4} + L^2_s -
\frac{\pi^2}{4}\right) +L_h (\gamma^0_j - \bar{\gamma}^0_h)
- 2 \bar{\gamma}^0_s L_s,
\\
B_4 &=& \frac{9}{2} C^2_F (\gamma^0_{\rm cusp})^2,
\\
B_3 &=& \frac{1}{2}C_F\gamma^0_{\rm cusp}\left(
9 C_F \gamma^0_{\rm cusp}L_h + 18 C_F \gamma^0_{\rm cusp}
L_s + 9\gamma^0_j - 18\bar{\gamma}^0_s - 5\beta_0 \right),
\\
B_2 &=& C_F \gamma^0_{\rm cusp} \left[ \frac{3
c^s_{11}}{C^2_A} + 3c^h_{11} + 3c^j_1 - L_h (3
\bar{\gamma}^0_h - 5\gamma^0_j + 4\bar{\gamma}^0_s +
\beta_0 ) + 4\gamma^0_j L_s - 14\bar{\gamma}^0_s L_s
\right.
\\
&&
\left.
- 4 L_s \beta_0 \right] + \frac{C_F ( (\Gamma^0_{h,21})^2 +
3C_A \gamma^1_{\rm cusp} )}{C_A} + 2\Gamma^0_{h,12}
\Gamma^0_{h,21} + \frac{1}{4} C^2_F (\gamma^0_{\rm cusp})^2
\nonumber
\\
&&
\times [ (L_h + 2L_s)(L_h + 14L_s) - 9 \pi^2 ] - \beta_0
(\gamma^0_j - 4\bar{\gamma}^0_s) -
4\gamma^0_j\bar{\gamma}^0_s +
(\gamma^0_j)^2+4(\bar{\gamma}^0_s)^2,
\end{eqnarray}
\begin{eqnarray}
B_1 &=& \frac{1}{12 C^2_A} C_F \gamma^0_{\rm cusp} \left[ 3
C^2_A [ 4 L_h (c^h_{11} - 2L_s (\bar{\gamma}^0_h -
\gamma^0_j + \bar{\gamma}^0_s)) + 8 L_s c^h_{11} + 4c^j_1
(L_h + 2 L_s)
\right.
\nonumber
\\
&&
+ L^2_h ( -4\bar{\gamma}^0_h + 3\gamma^0_j
+2\bar{\gamma}^0_s ) + 4\gamma^0_j L^2_s - 3\pi^2
\gamma^0_j - 24\bar{\gamma}^0_s L^2_s + 6\pi^2
\bar{\gamma}^0_s )]
\nonumber
\\
&&\left.
+ 12 c^s_{11} (L_h + 2 L_s) + C^2_A
\beta_0 (5\pi^2 - 6(L^2_h+4L^2_s))\right]
\nonumber
\\
&&
+ C_F \left[
\frac{\Gamma^0_{h,21}}{2C_A}(2c^h_{12}-\Gamma^0_{h,21}
(L_h+2L_s)) + \gamma^1_{\rm cusp} (L_h+2L_s) \right]
\nonumber
\\
&&+ \frac{1}{C^2_A} (2\Gamma^0_{h,21} c^s_{12} + c^s_{11}
(\gamma^0_j - 2(\bar{\gamma}^0_s +
\beta_0)))+2\Gamma^0_{h,12}c^h_{12} +(2L_s-L_h)
\Gamma^0_{h,12} \Gamma^0_{h,21}
\nonumber
\\
&&
+ c^h_{11}(\gamma^0_j - 2 \bar{\gamma}^0_s ) -
\frac{C^2_F}{4} (\gamma^0_{\rm cusp})^2[(L_h+2
L_s)(L^2_h-4L^2_s+3\pi^2)-36\zeta_3]
\nonumber
\\
&&
+ c^j_1 (\gamma^0_j - 2 \bar{\gamma}^0_s - \beta_0 )
 - \bar{\gamma}^0_h \gamma^0_j L_h + 2\bar{\gamma}^0_h
\bar{\gamma}^0_s L_h - 2 \gamma^0_j \bar{\gamma}^0_s L_h -
2 \gamma^0_j \bar{\gamma}^0_s L_s - \gamma^0_j \beta_0 L_h
\nonumber
\\
&&
+ (\gamma^0_j)^2 L_h + \gamma^1_j + 4 \bar{\gamma}^0_s
\beta_0 L_s + 4(\bar{\gamma}^0_s)^2 L_s - 2\bar{\gamma}^1_s,
\end{eqnarray}
where $L_h = \ln\frac{\mtsq}{\mu^2}$ and
$L_s=\ln\frac{\mtsq \sqrt{\sss}}{(\sss-\mtsq)\mu}
$. Explicit expressions for $c^h$, $c^s$
and $c^j$ are given in the appendix. Several comments on
this result are in order. Our NNLO expansion are
accurate to NNLL accuracy, \emph{i.e.}, the coefficients
$A_{2,1,0}$ and $B_{4,3,2,1}$ are accurate. Complete
expression for $B_0$ can only be known from a 2-loop
calculation, therefore we retain from giving a partial
result for it here. Similar expansion has been performed
in the Ref.~\cite{Kidonakis:2010tc}, but only partial NNLL
logarithms are presented. In particular, the coefficient
of $D_2$ and $D_1$ at NNLO is not complete
in~\cite{Kidonakis:2010tc}, because of the omission
of virtual corrections in their work. We also note that due
to different definition of $s_4$, in order to comparison
with the result in the Ref.~\cite{Kidonakis:2010tc}, we have
to make the follwing change to $\bar{\gamma}_s$
\begin{equation}
\bar{\gamma}_s = -C_F \gamma_{\rm cusp}
\ln\frac{(\sss-\mtsq)^2\sqrt{\sss}}{\uuu\ttt m_t} + \gamma_s
\rightarrow  -C_F \gamma_{\rm cusp}
\ln\frac{(\sss-\mtsq)^2\sqrt{\sss}}{(\uuu-\mtsq)(\ttt-\mtsq)
m_t} + \gamma_s.
\end{equation}
We have checked that our expression for $B_{4,3}$ agree with
the Ref.~\cite{Kidonakis:2010tc} after this replacement.
We also checked that our $\mu$ dependence in the expansion
coefficients agree with \cite{Kidonakis:2010tc} for
$A_{2,1}$ and $B_{4,3,2}$ after this replacement. There is
an extra $\mu$ dependence term in $B_4$ of our
expansion, $C_F \gamma^0_{\rm cusp}L_h c^h_{11}$, which is
not presented in \cite{Kidonakis:2010tc}. There is also
$\mu$ dependence in $B_0$. However we are not able to
check this term against \cite{Kidonakis:2010tc} since the
explicit expression for this term is not presented there.
We give a numerical comparison on the difference of
different threshold variable definition and the effects of
omitting virtual corrections in the resummation result in
next section.

\section{Numerical discussion}
\label{sec:6}

In this section, we present the numerical results for the
threshold
resummation effects on the s-channel single top production
at both
the Tevatron and the LHC. The input parameters used
throughout this
section are given below:
\[
\begin{gathered}
 m_t=173.2\;{\rm{GeV}},\quad M_W=80.4\;{\rm{GeV}},\quad
G_F=1.16639\times 10^{-5}\;{\rm{GeV}}, \\
V_{\rm CKM} = \left(
\begin{tabular}{ccc}
0.9751 & 0.2210 & 0\\
0.2215 & 0.9743 & 0\\
0.0035 & 0.0410 & 1
\end{tabular}
\right).
\end{gathered}
\]
We use the MSTW2008NNLO PDFs~\cite{Martin:2009iq}
throughout our numerical calculation.

\FIGURE{
\centering
\begin{tabular}{ccc}
\epsfig{file=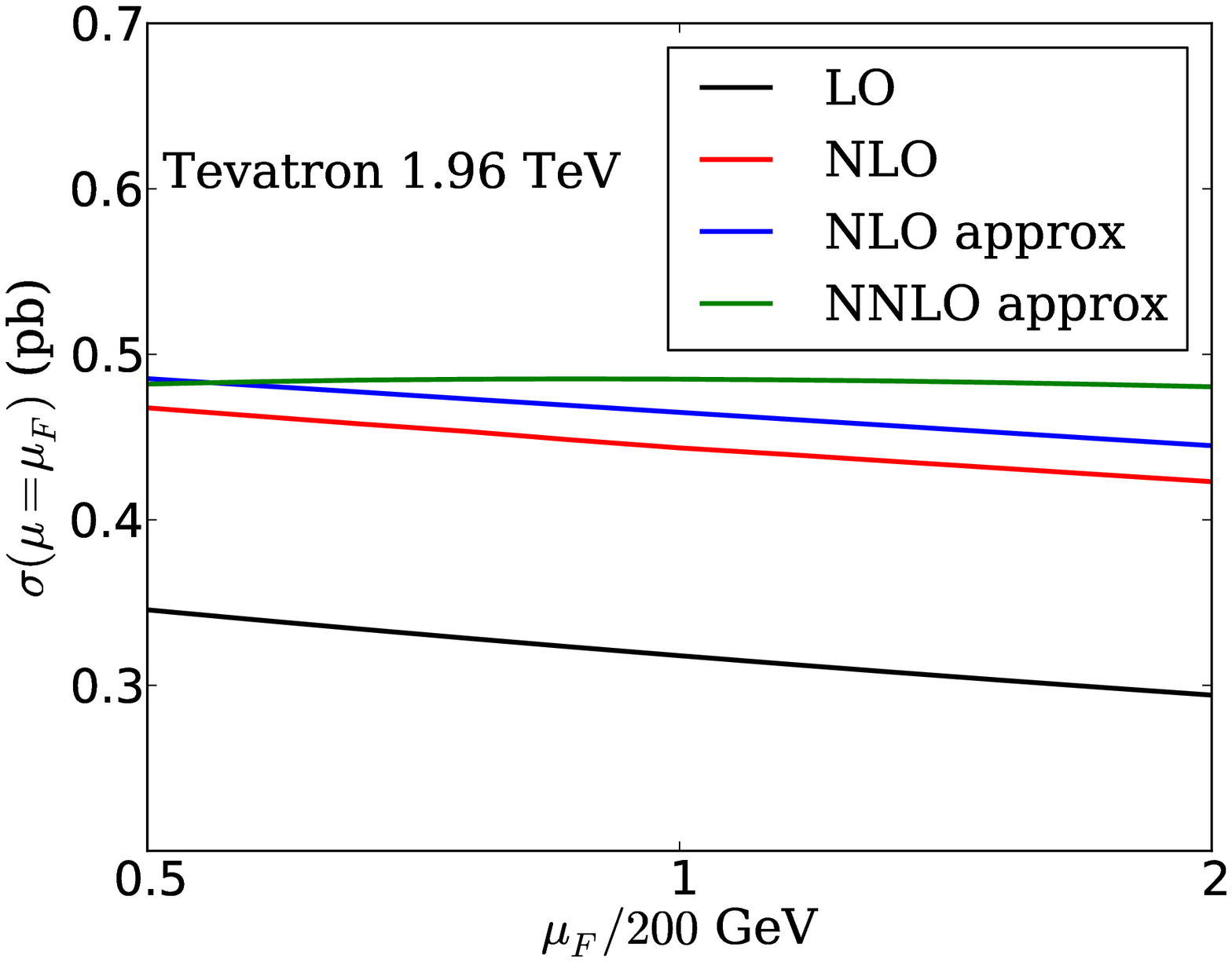,width=0.3\linewidth,
clip=}
&
\epsfig{file=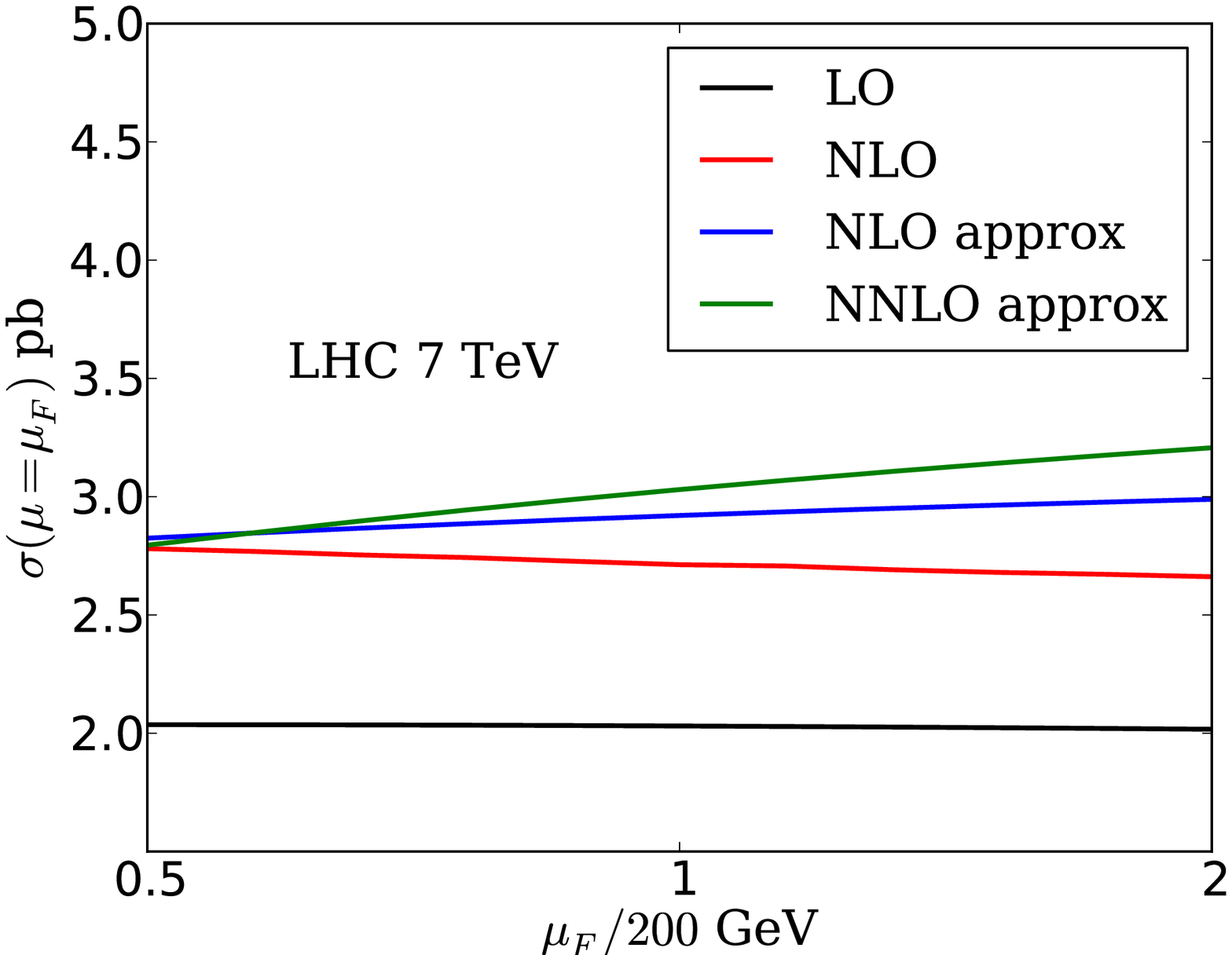,width=0.3\linewidth,
clip= }
&
\epsfig{file=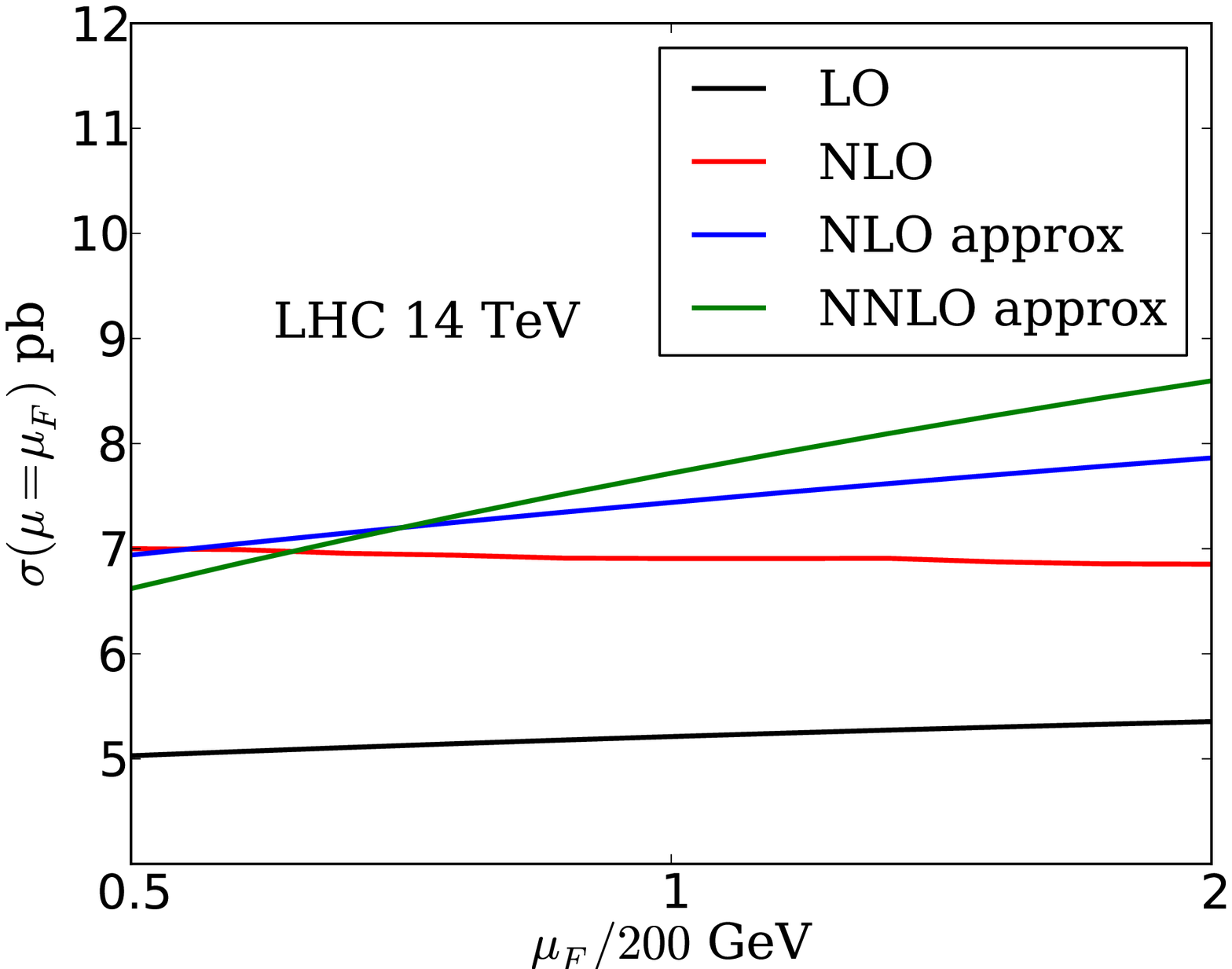,width=0.3\linewidth,
clip= }
\end{tabular}
\caption{Factorization scale dependence of NNLO expanded
cross sections for s-channel single
top production at the Tevatron ($1.96$ TeV) and LHC ($7$
TeV and $14$ TeV).}
\label{scale_v}
}

Before presenting the numerical results for resummed cross section,
it's important to examine to what extent the singular terms
approximate the fixed order calculation. It would be meaningless if
the neglected subleading terms are as important as the singular
terms. We present the numerical results for the approximate NLO and
NNLO cross section, Eq.~(\ref{nnlo_expanded}), for both Tevatron and
LHC in Fig.~\ref{scale_v}. The factorization scale is chosen as
$\mu_F=200$ GeV, as will be explained below. It can be seen from
Fig.~\ref{scale_v} that at the Tevatron, which has a lower collision
energy, the approximate NLO cross section over estimates the exact
NLO cross section by about $5\%$. Since the NLO corrections for
s-channel single top production is quite large, about $40\%$ at the
Tevatron, we consider the threshold expansion as a good
approximation. Furthermore, the fact that the scale dependence of
the approximate NLO result is similar to exact NLO result implies
the small scale dependence of the subleading terms. On the other
hand, the NLO approximation doesn't work well at the LHC with higher
collision  energy at $7$ TeV or $14$ TeV, as shown in
Fig.~\ref{scale_v}. The differences mentioned above are smaller at
lower factorization scale and more significant at larger
factorization scale. Moreover, the scale dependence of the
approximate NLO cross section behave quite differently from the
exact NLO results, indicating that the subleading terms are not only
numerically large, but also have large impact on the factorization
scale dependence of the cross section.
Therefore, we conclude that our resummation results at the
LHC are not as reliable as at the Tevatron. Nevertheless,
we still give the NNLO approximate and resummed
results for both the Tevatron and the LHC as a reference.

In order to calculate the resummed cross section,
Eq.~(\ref{diffxec}), we need to
determine the
appropriate scales for the process. In the SCET approach to
resummation, there are four scales: the hard scale $\mu_h$,
the jet
scale $\mu_j$, the soft scale $\mu_s$, and the factorization
scale
$\mu_F$, and we have assumed that the renormalization scale
equals
the factorization scale for simplicity. This is different
from the
fixed order calculation, where only the factorization scale
is
accessible. This is actually the merit of effective theory,
since
the calculation has been factorized into a series of
single-scale
problems, and large logarithms can be avoided if appropriate
value
of scale is chosen for each problem.

First, we choose a default value for the hard scale. Since
s-channel
single top production is similar to the Drell-Yan process,
one would
expect that the appropriate value of the hard scale should
be around
$\sqrt{\hat{s}}$. However, choosing
$\mu_h\sim\sqrt{\hat{s}}$ is
inconvenient because $\sqrt{\hat{s}}$ is a dynamical
variable. To
avoid this inconvenience we set the hard scale at a fixed
value,
$\mu_h=200$ GeV, which is slightly larger than $m_t$. We
have
checked that invariant mass distribution of the virtual $W$
boson
peaks around this value, and thus it can be considered as
the
``average'' value of $\sqrt{\hat{s}}$. For simplicity, we
also
choose the factorization scale to be $200$ GeV.

Next, we determine the appropriate soft scale and jet scale.
In
order to obtain a reasonable physical result, we expect that
no
large logarithms should arise in the soft and jet function
when the
appropriate scales are chosen. In practice, for the
determination of
soft scale (jet scale), we fix the factorization scale at
$200$ GeV
in the factorized cross section Eq.~(\ref{diffxec}), and set
all the
other scales equal to $\mu_s$ ($\mu_j$), and then vary
$\mu_s$
($\mu_j$). In Fig.~\ref{sj}, we plot the cross section which
only
include the one-loop soft corrections or jet corrections,
respectively, divided by the tree-level cross section with
all
scales equal to $200$ GeV. From Fig.~\ref{sj}, we find that
the net
corrections are small around $\mu_j=50$ GeV for the jet
corrections. Thus we
take it as our default scale choice for jet scale. The
choice of soft scale is not as clear as jet scale from
Fig.~\ref{sj} since the total soft corrections do not have
a clear minimum or maximum. The reason is that at NLO, the
soft corrections consist of a Drell-Yan like corrections in
the initial state and a soft gluon corrections in the final
state, as depicted in graphs (a), (b) and (c) of
Fig.~\ref{softint}. It turns out that the corrections is
positive in the initial state and negative in the final
state. Either initial state corrections or final state
corrections show an appropriate scale around $25$ GeV, but
not the sum. On the other hand, an appropriate choice for
soft scale is dictated by the picture of underlying
factorization, $\mu_s\sim\mu^2_j/\mu_h$. From our choice for
hard and jet scale, this implies that soft scale should be
chosen around $10-30$ GeV. In our numerical calculation, we
have chosen $\mu_s$ as $25$ GeV, to avoid too close to
$\Lambda_{\rm QCD}$ while probe as much soft activity as
possible.
\FIGURE{
 \centering
 \parbox{2.5in}{
 \includegraphics[width=2.5in]{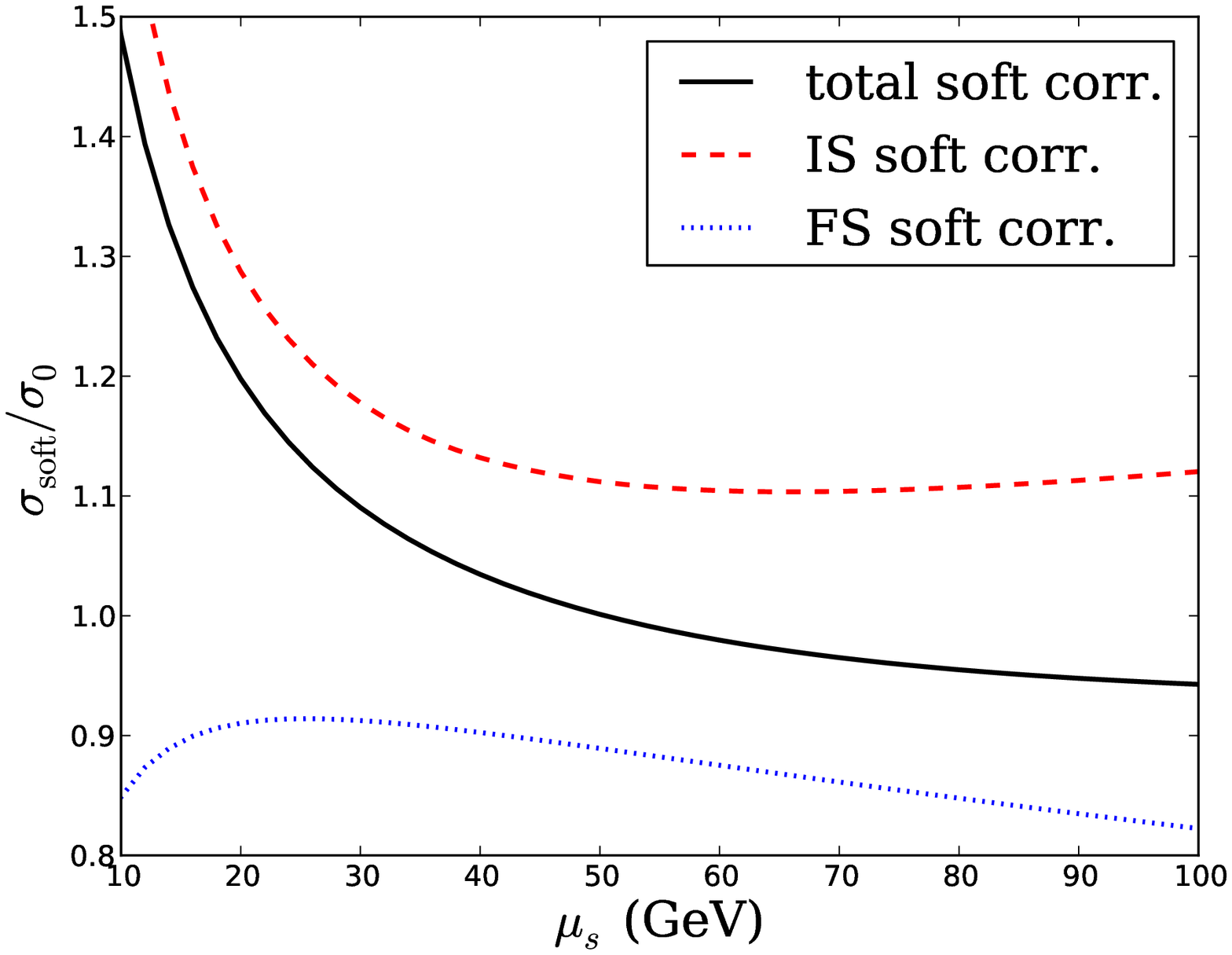}}
\qquad
\parbox{2.5in}{
 \includegraphics[width=2.5in]{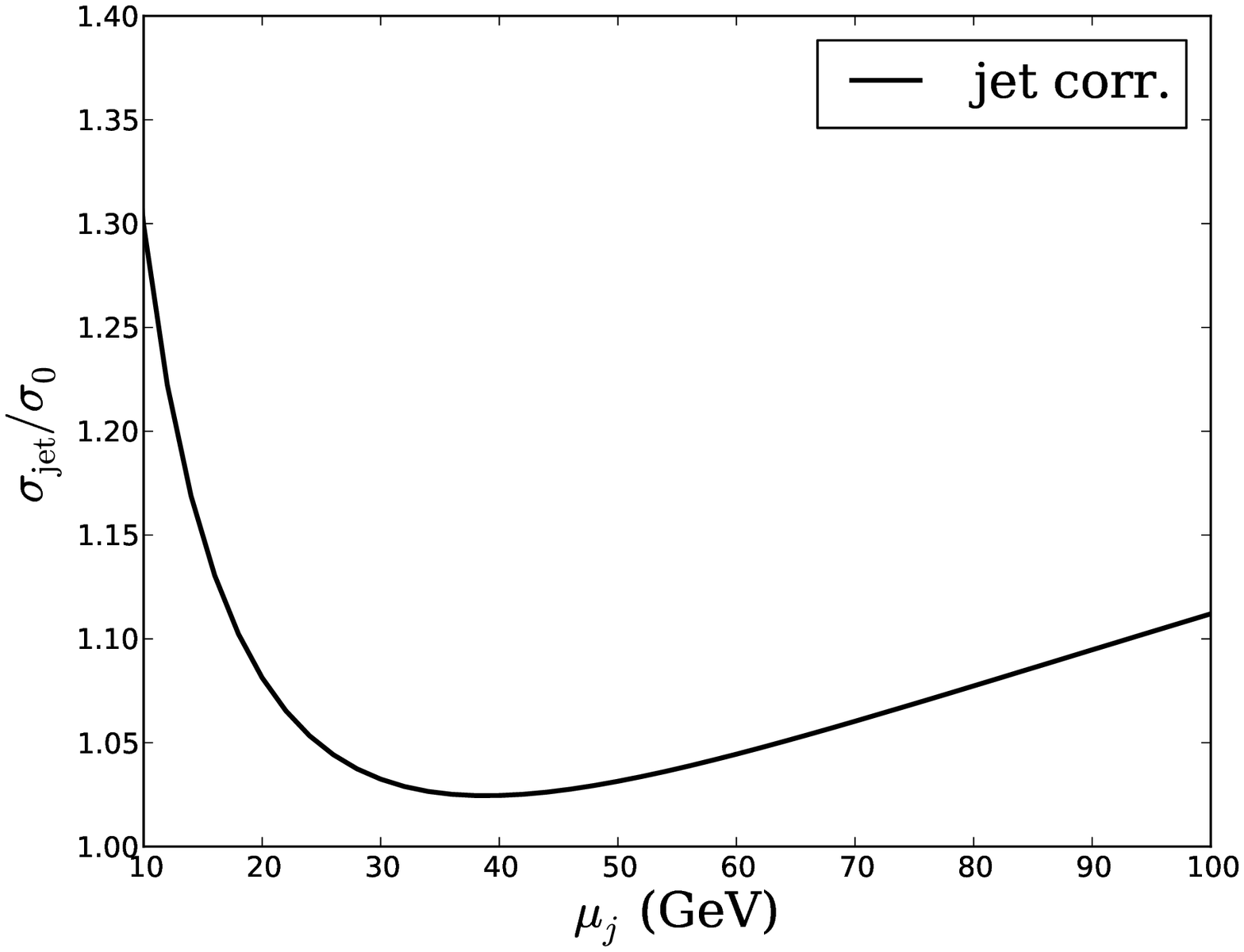}
}
\caption{Determination of $\mu_s$ and $\mu_j$. On the left
(right) is the cross section with only soft
(jet) corrections, divided by the tree-level cross section,
for the variation of $\mu_s$ ($\mu_j$). The red dashed line
includes only the initial state soft corrections, while the
blue dotted line includes only the final state soft
corrections. The
factorization scale is fixed at $200$ GeV.}
\label{sj}
}

We also note that the factorized cross section
Eq.~(\ref{diffxec})
is derived in the threshold limit, $s_4\to 0$. To capture
the
non-leading terms, we must match the resummed cross section
onto the
NLO cross section, which can be found in the
Ref.~\cite{Harris:2002md}. We have redone the calculation
and found
complete agreement with the Ref.~\cite{Harris:2002md}. After
matching in momentum space approach, the resummed total
cross
sections is given by
\begin{equation}
 \sigma^{\rm RES}=\sigma^{\rm thres} - \left.\sigma^{\rm
thres}\right|_{\mu_h=\mu_j=\mu_s=\mu_F} + \left. \sigma^{\rm
NLO}\right|_{\mu_F}.
\end{equation}

In table~\ref{tab1}, we present the resummed total cross
section for
s-channel single top and anti-top production, as well as the
NNLO approximation at both the Tevatron
and the LHC. All the
scales are
set to the default values, \emph{i.e.}, $\mu_F=\mu_h=200$
GeV,
$\mu_s=25$
GeV and $\mu_j=50$ GeV. It can be seen that the NLO QCD
corrections
significantly enhance the total cross section at both the
Tevatron
and the LHC~\cite{Harris:2002md}. The threshold resummation
effects
further increase the NLO cross section by about $3\%-5\%$
at the Tevatron.
However,
we do not observe the large enhancement due to threshold
resummation
as reported in Ref.~\cite{Kidonakis:2010tc}. {The
discrepancy has two origin. First, the 1-loop matching
coefficients
of hard function was not taken into account
in~\cite{Kidonakis:2010tc}. second, the definiton of $s_4$
used
in~\cite{Kidonakis:2010tc} does not coincide with ours. As
explained in Sec.~\ref{sec:3}, the definition we use also
include
the effects of collinear splitting of final state b-quark,
and
should be considered as a better choice.} We quantify the
numerical significance of the difference from these
different treatment in the end of this section briefly. We
also show the resummed cross section for single top
production at the Tevatron for different top quark mass in
Fig.~\ref{mt_var}. It can be seen that the LO prediction
significantly under estimates the total cross section. It's
also clear that the resummed cross section dramatically
improves the scale dependence, comparing with the NLO
results.
\TABLE{
\begin{tabular}{ccccc}
\hline
\hline
   & $\sigma_{\rm LO}$ &$\sigma_{\rm
NLO}$& $\sigma_{\rm expand}$ &  $\sigma_{\rm
RES}$\\
\hline
Tevatron (top) & $0.318^{+0.029}_{-0.024}$ pb &
$0.443^{+0.024}_{-0.020}$ pb&  $0.463^{+0.002}_{-0.004}$
pb &  $0.467^{+0.010}_{-0.010}$ pb\\
\hline
Tevatron (anti-top)  & $0.318^{+0.029}_{-0.024}$ pb &
$0.443^{+0.024}_{-0.020}$ pb&$0.463^{+0.002}_{-0.004}$
pb&  $0.467^{+0.010}_{-0.010}$
pb\\
\hline
LHC ($7$ TeV, top) & $2.03^{+0.01}_{-0.01}$ pb &
$2.71^{+0.07}_{-0.05}$ pb & $2.82^{+0.06}_{-0.07}$ pb  &
$2.81^{+0.16}_{-0.10}$ pb \\
\hline
LHC ($7$ TeV, anti-top) & $1.14^{+0.01}_{-0.01}$ pb &
$1.53^{+0.04}_{-0.03}$ pb & $1.60^{+0.03}_{-0.04}$ pb  &
$1.60^{+0.08}_{-0.05}$ pb\\
\hline
LHC ($14$ TeV, top) & $5.21^{+0.14}_{-0.18}$ pb&
$6.91^{+0.09}_{-0.05}$ pb& $7.17^{+0.20}_{-0.25}$ pb  &
$7.11^{+0.47}_{-0.35}$ pb
\\
\hline
LHC ($14$ TeV, anti-top) & $3.36^{+0.09}_{-0.12}$ pb &
$4.46^{+0.03}_{-0.05}$ pb &$4.64^{+0.10}_{-0.18}$ pb &
$4.61^{+0.28}_{-0.24}$ pb
\\
\hline
\end{tabular}
\caption{Total cross section for single top and
anti-top production at the Tevatron and LHC. All the scales
are chosen at the default values. For the resummed
results, The total uncertainties
are obtained by adding the individual scale
variations of $\mu_F,\mu_h,\mu_j,\mu_s$ in quadrature.}
\label{tab1}
}

\FIGURE{
 \centering
 \includegraphics[width=4in]{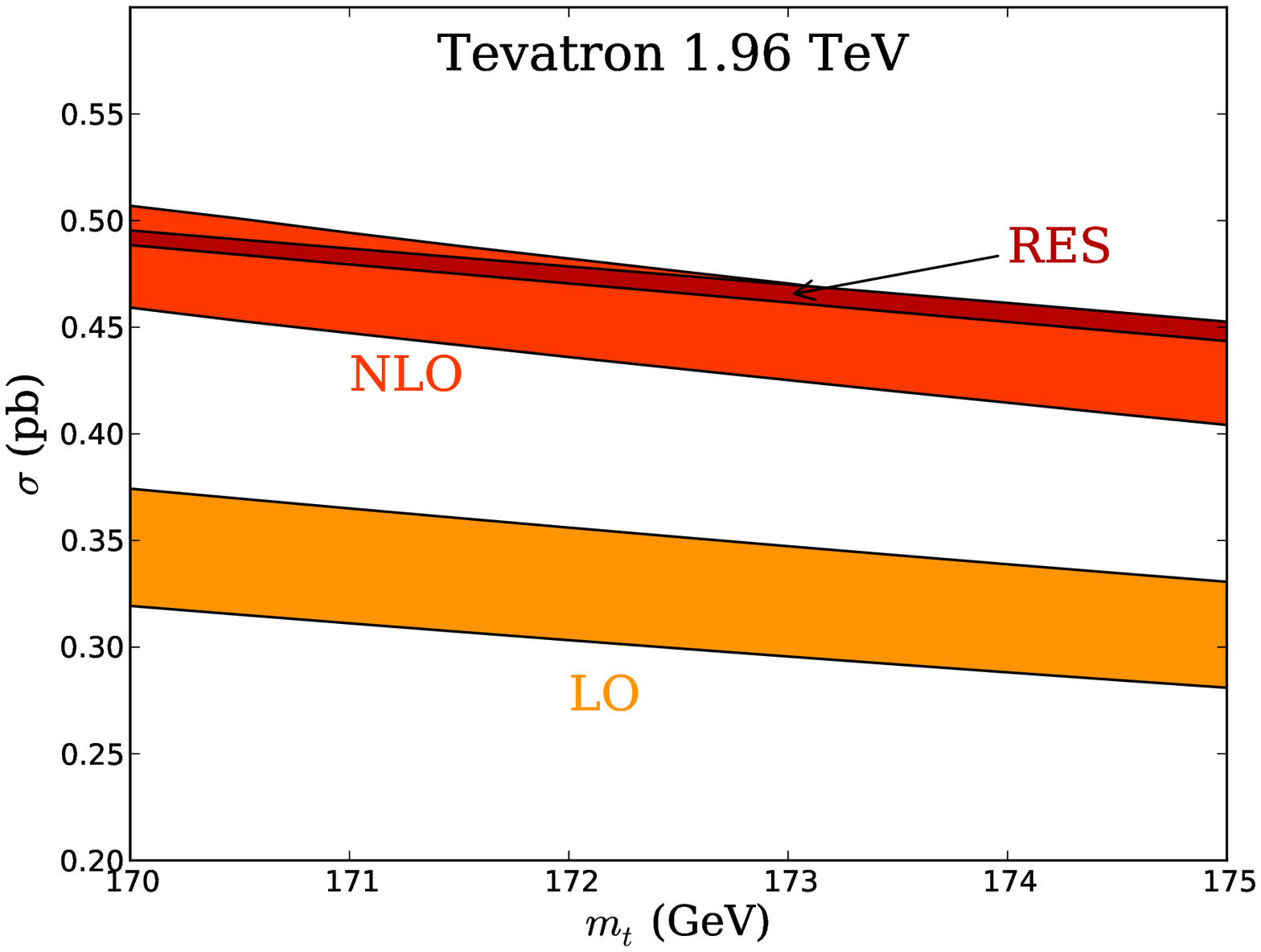}
\caption{Top quark mass dependence of the fixed order and
resummed cross section. The band corresponds to the
variation of factorization scale around $200$ GeV by a
factor of 2.}
\label{mt_var}
}

Finally, we give a brief numerical comparision with
results
presented in Ref.~\cite{Kidonakis:2010tc}. As mentioned in
the last section, our NNLO singular
expansion differ from those presented
in~\cite{Kidonakis:2010tc} in two aspects: (a) We have
included all the 1-loop matching coeffcients $c^h,c^s,c^j$
in our calculation. Therefore our results contain all the
NNLL logarithms.~\footnote{We refer to
Ref.~\cite{Becher:2007ty} for the accurate definition of
logarithmic order in SCET approach to resummation.} (b) The
results in the Ref.~\cite{Kidonakis:2010tc} used a
different definition of $s_4$, as was mentioned in
Sec.~\ref{sec:3}. To quantify the effects of these
differences, we plot two sets of cross sections below.
In Fig.~\ref{virt}, we plot the NNLO approximate result
with $c^h_{ij}$ set to zero. As can be seen from the
figure, the NLO or NNLO approximation fail to
approximate the exact NLO result when the virtual
corrections are off, which give large positive contribution
at the NLO. In Fig.~\ref{s4def}, we plot the NNLO
approximate results with the same definition of $s_4$ as in
Ref.~\cite{Kidonakis:2010tc}. The NLO or NNLO approximate
cross sections are significantly enhanced with such choice
for $s_4$, and overestimate the exact NLO results by a large
amount.

\FIGURE{
\centering
\begin{tabular}{ccc}
\epsfig{file=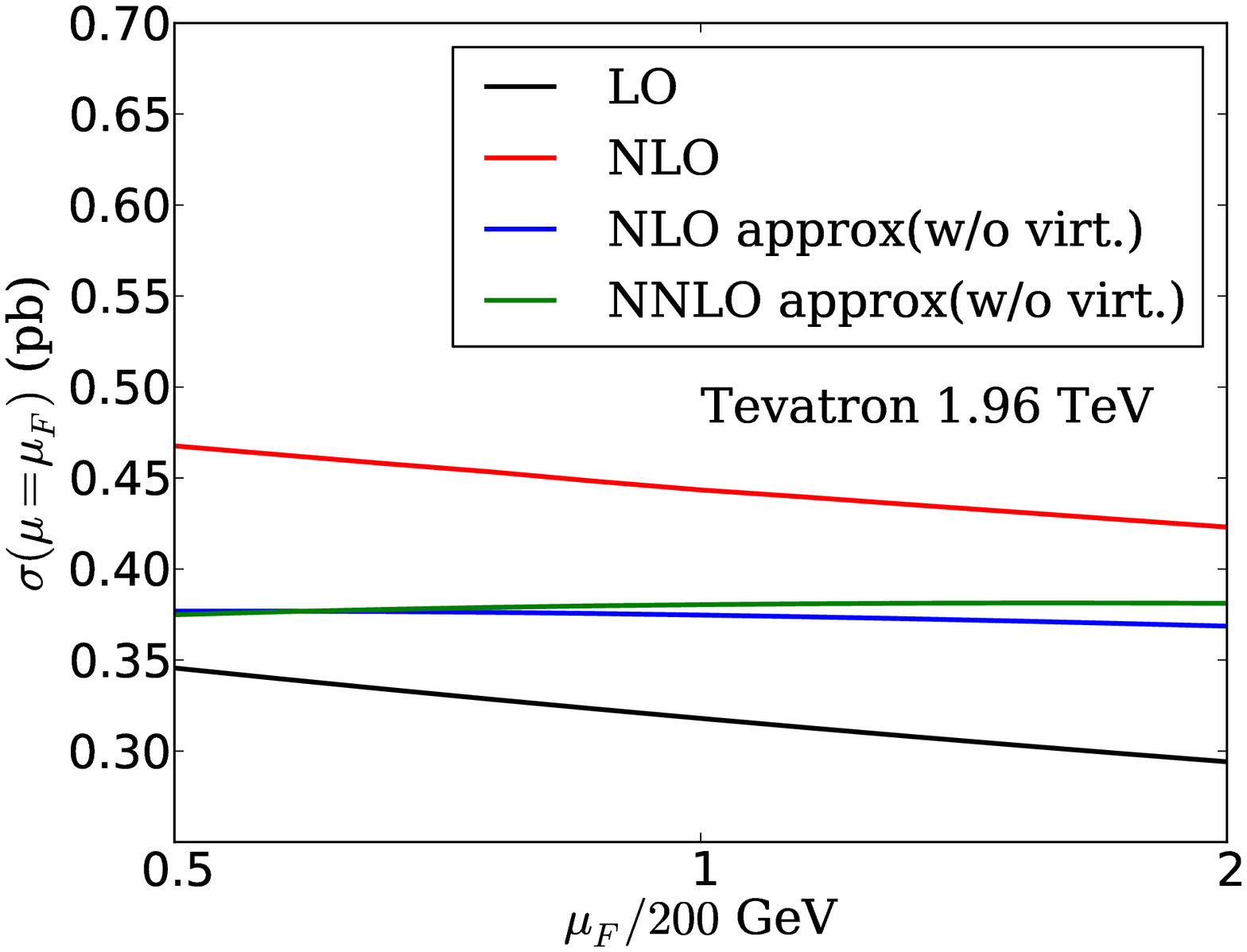,width=0.3\linewidth,
clip=}
&
\epsfig{file=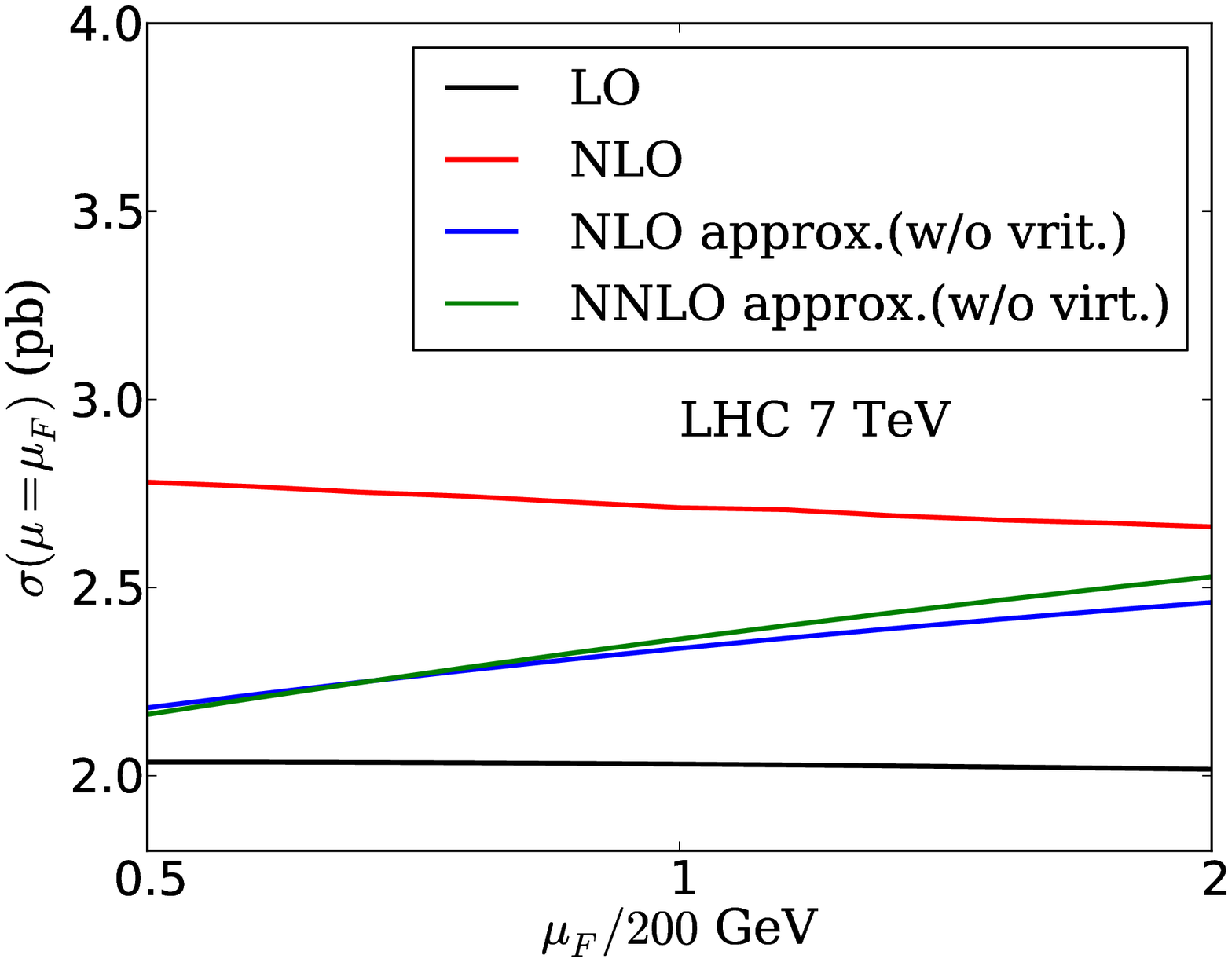,width=0.3\linewidth,
clip= }
&
\epsfig{file=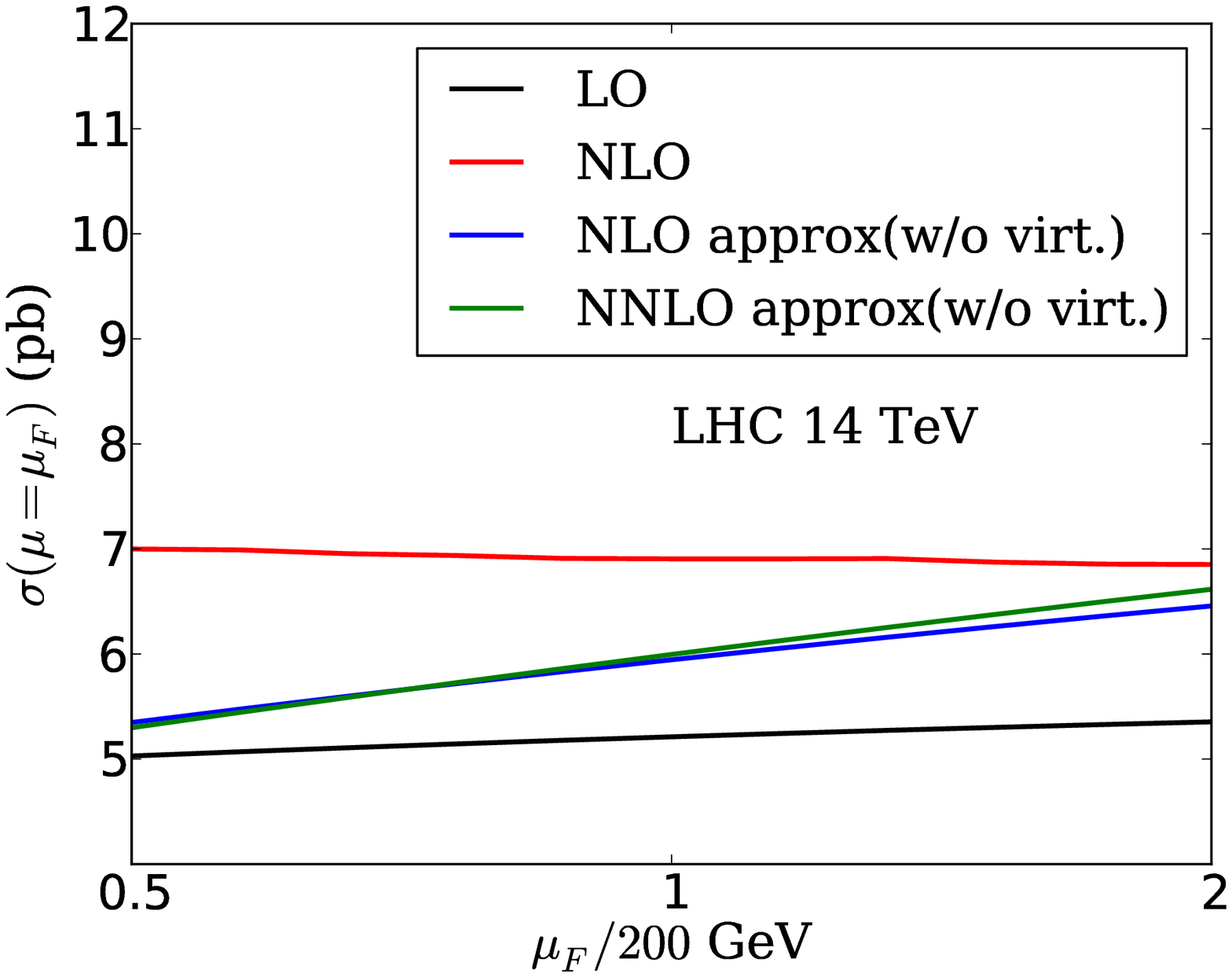,width=0.3\linewidth,
clip= }
\end{tabular}
\caption{Illustration of the importance of the NLO virtual
corrections in the NNLO expansion. We have set $c^h_{ij}$
to zero in these plots.}
\label{virt}
}

\FIGURE{
\centering
\begin{tabular}{ccc}
\epsfig{file=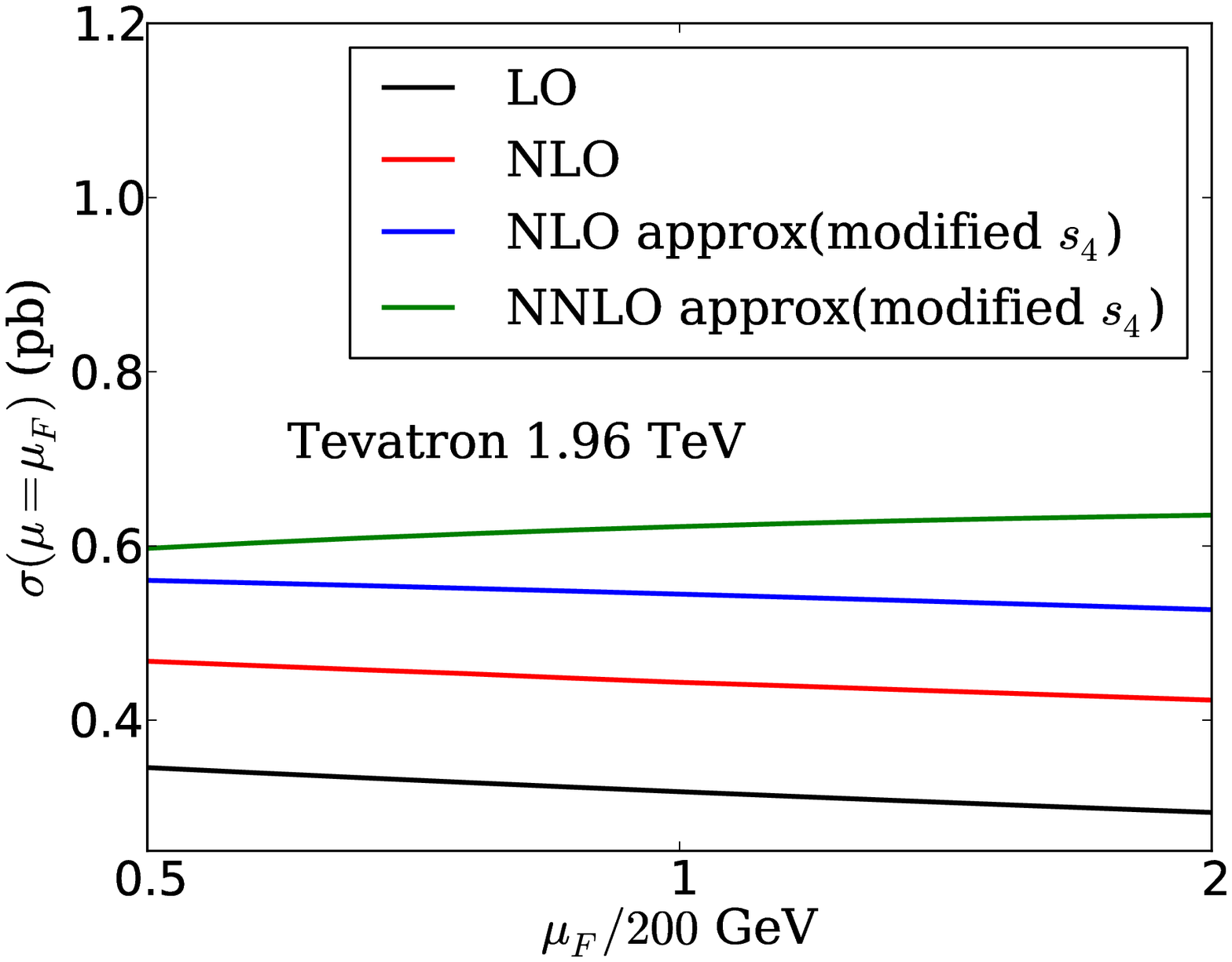,width=0.3\linewidth,
clip=}
&
\epsfig{file=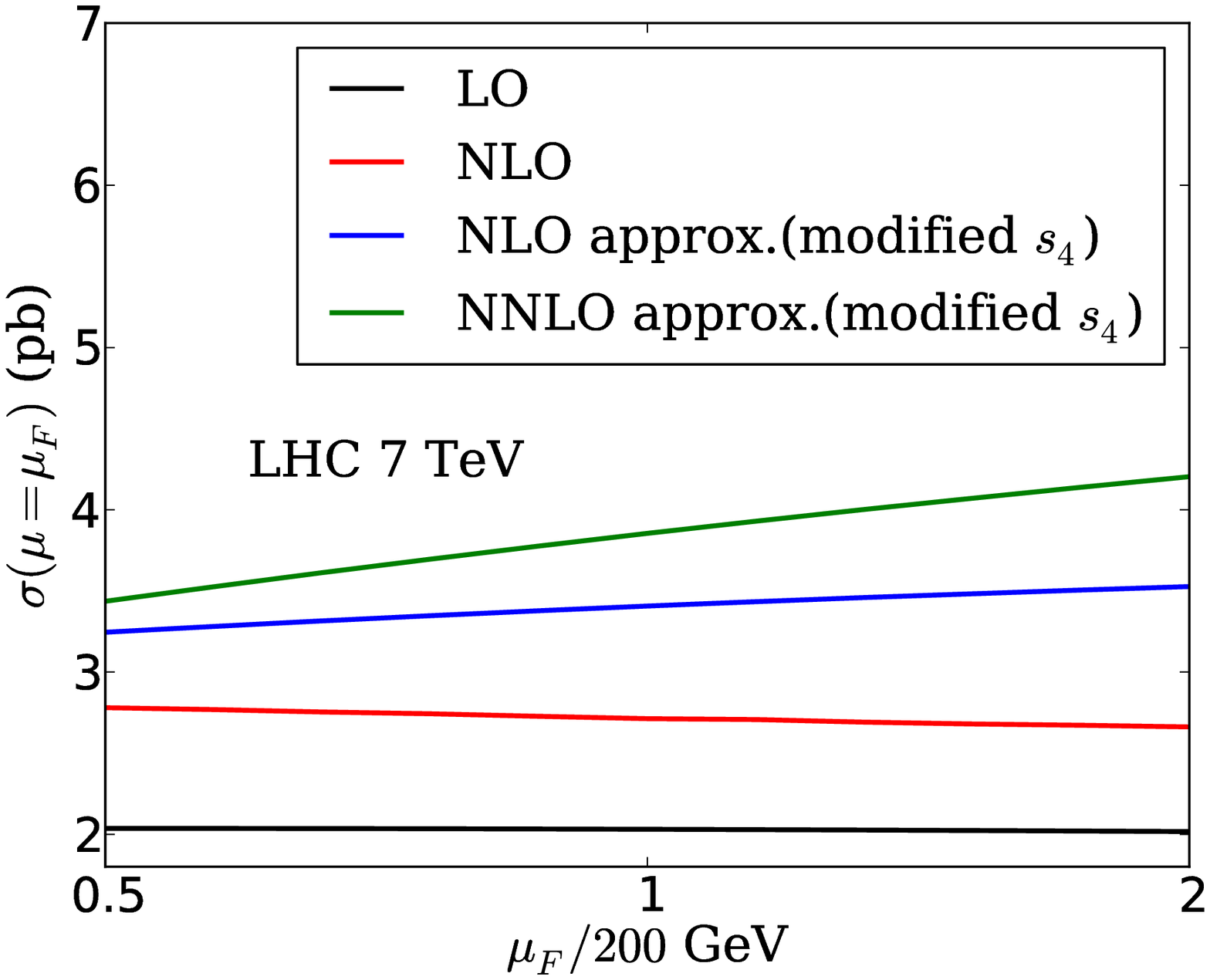,width=0.3\linewidth,
clip= }
&
\epsfig{file=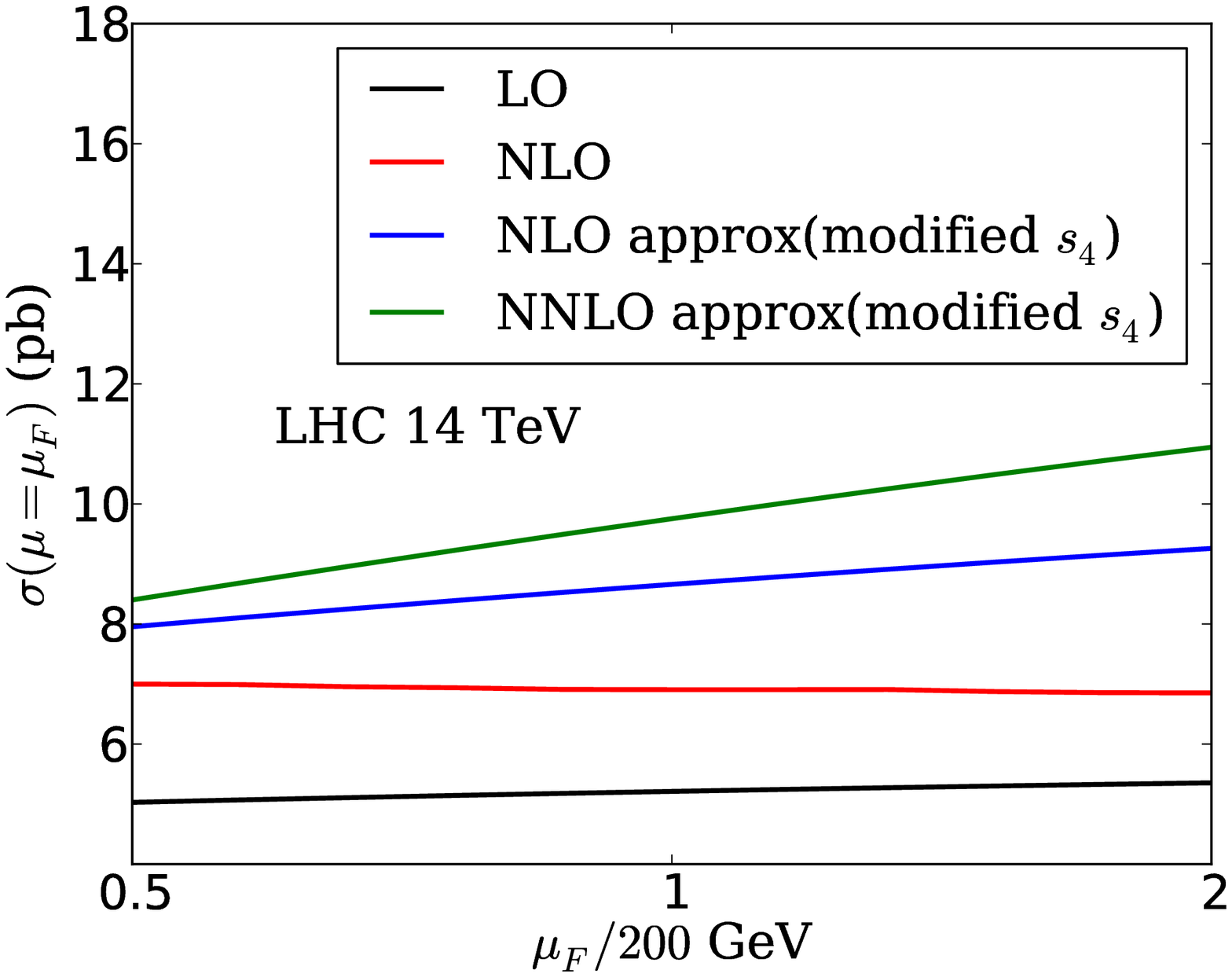,width=0.3\linewidth,
clip= }
\end{tabular}
\caption{Illustration of the effects of different threshold
variable definition. We have chosen the same $s_4$
definition as Ref.~\cite{Kidonakis:2010tc} in these plots.}
\label{s4def}
}

\section{Conclusions}
\label{sec:7}

We have studied the production of s-channel single top quark
in the
SM at both the Tevatron and the LHC. Using SCET, we show
that the
production cross section can be factorized into a
convolution of
hard function, soft function and jet function in the
threshold
limit. Each function, being sensitive to a single scale, is
free of
large logarithms once an appropriate scale is chosen. By
this way,
the threshold resummation is performed with the conventional
RG
equation. As a by-product, we obtain a NNLO expansion of
threshold
singular distributions. We also perform a numerical
investigation of
our resummed formula, using the momentum space resummation
formalism~\cite{Becher:2006nr}. We find that in general, the
higher order threshold logarithms enhance the NLO cross
sections by about $3\%-5\%$ at the Tevatron, and the
resummation effects significantly reduce the factorization
scale dependence of the total cross section at the Tevatron,
while at the LHC the factorization scale dependence has not
been improved, compared with the NLO results.

\acknowledgments We would like to thank Matthew Schwartz and
Li Lin
Yang for useful discussions. This work is supported in part
by the
National Natural Science Foundation of China, under Grants
No.~11021092 and No.~10975004.

%%%%%%%%%%%%%%%%%%%%%%%%%%%%%%%%%%%%%%%%%%%%%%%%%%%%%%%%%%%%
%%%%%%%%%%
%%%%

\appendix
\section{Relevant anomalous dimensions and matching
coefficients}
The various anomalous dimensions
needed in our resummation can be found, \emph{e.g.}, in the
Refs~\cite{Becher:2006mr,Becher:2007ty,Becher:2009th}. We
list them below for the convenience of the reader. The QCD
$\beta$
function is
\begin{equation}
 \beta (\alpha_s) = -2 \alpha_s \left[ \beta_0
\frac{\alpha_s}{4\pi} + \beta_1 \left(
\frac{\alpha_s}{4\pi} \right)^2 + \cdots \right],
\end{equation}
with expansion coefficients
\begin{eqnarray}
 \beta_0 &=& \frac{11}{3} C_A - \frac{4}{3} T_F n_f,
\nn
\\
\beta_1 &=& \frac{34}{3}C^2_A - \frac{20}{3} C_A T_F n_f -
4C_F T_F n_f,
\nn
\\
\beta_2 &=& \frac{2857}{54} C^3_A + \left( 2 C^2_F -
\frac{205}{9} C_F C_A - \frac{1415}{27} C^2_A \right) T_F
n_f + \left( \frac{44}{9}C_F + \frac{158}{27} C_A \right)
T^2_F n^2_f,
\end{eqnarray}
where $C_A=3$, $T_F=1/2$ for QCD, and $n_f$ is
the number of active quark flavor.

The cusp anomalous dimension is
\begin{equation}
\label{cuspa}
 \gamma_{\rm cusp} (\alpha_s) = \gamma^0_{\rm cusp}
\frac{\alpha_s}{4\pi} +
\gamma^1_{\rm cusp} \left(
\frac{\alpha_s}{4\pi} \right)^2 + \cdots,
\end{equation}
with
\begin{eqnarray}
 \gamma^0_{\rm cusp} &=& 4,
\nn
\\
\gamma^1_{\rm cusp} &=& 4 \left[ \left( \frac{67}{9} -
\frac{\pi^2}{3} \right) C_A - \frac{20}{9} T_F n_f \right],
\nn
\\
\gamma^2_{\rm cusp} &=& 4 \left[ C^2_A \left(
\frac{245}{6} -
\frac{134}{27}\pi^2 + \frac{11}{45}\pi^4 +
\frac{22}{3}\zeta_3 \right) + C_A T_F n_f \left(
-\frac{418}{27} + \frac{40}{27}\pi^2 - \frac{56}{3}\zeta_3
\right)
\right.
\nn
\\
&&\left. + C_F T_F n_f \left( -\frac{55}{3} + 16 \zeta_3
\right) - \frac{16}{27} T^2_F n^2_f \right].
\end{eqnarray}

The other anomalous dimensions are expanded as
Eq.~(\ref{cuspa}), and their expansion coefficients are
\begin{eqnarray}
 \gamma^{0}_q &=& -3C_F,
\nn
\\
\gamma^{1}_q &=& C^2_F\left(-\frac{3}{2}+2\pi^2-24
\zeta_3\right) + C_F
C_A \left( -\frac{961}{54}-\frac{11}{6}\pi^2 + 26\zeta_3
\right) + C_F T_F n_f \left( \frac{130}{27} +
\frac{2}{3}\pi^2 \right),
\nn
\\
\gamma^{0}_Q &=& -2C_F,
\nn
\\
\gamma^{1}_Q &=& C_F C_A \left( \frac{2}{3}\pi^2 -
\frac{98}{9} - 4 \zeta_3 \right) + \frac{40}{9} C_F T_F n_f,
\nn
\\
\gamma^{0}_\phi &=& 3C_F,
\nn
\\
\gamma^{1}_\phi &=& C^2_F \left( \frac{3}{2} - 2 \pi^2 +
24 \zeta_3 \right) + C_F C_A \left( \frac{17}{6} +
\frac{22}{9} \pi^2 - 12 \zeta_3 \right) - C_F T_F n_f
\left( \frac{2}{3} + \frac{8}{9} \pi^2 \right),
\nn
\\
\gamma^{0}_j &=& -3C_F,
\nn
\\
\gamma^{1}_j &=& C^2_F \left( -\frac{3}{2} + 2 \pi^2-
24 \zeta_3 \right) + C_F C_A \left( -\frac{1769}{54} -
\frac{11}{9} \pi^2 + 40 \zeta_3 \right)
\nn
\\
&& + C_F T_F n_f
\left( \frac{242}{27} + \frac{4}{9} \pi^2 \right),
\end{eqnarray}
$\gamma_h$ and $\gamma_s$ can be obtained from the
anomalous dimensions above through the following equations:
\begin{eqnarray}
  \gamma_{h} &=& 3 \gamma_q + \gamma_Q,
\nn
\\
\gamma_{s} &=& -2 \gamma_\phi - \gamma_{h} + \gamma_j.
\end{eqnarray}

The hard function is a $2\times 2$ matrix in
color space.
To $\mathcal{O}(\alpha_s)$, it can be written as
\begin{equation}
 \textbf{H} = \left(
\begin{array}{cc}
1 + \frac{\alpha_s}{4\pi} H^1_{11} &
\frac{\alpha_s}{4\pi} H^1_{12}\\
\frac{\alpha_s}{4\pi} H^1_{21} & 0
\end{array}\right).
\end{equation}
$H^1_{11}$ can be obtained from evaluating the
first two diagrams of Fig.~\ref{box} and the corresponding
counter-terms. It is given by
\begin{eqnarray}
  H^1_{11} = -\frac{3}{4} C_F \gamma^0_{\rm cusp}(\alpha_s)
\ln^2\frac{m^2_t}{\mu^2} -
\bar{\gamma}^0_h\ln\frac{m^2_t}{\mu^2} + c^h_{11},
\end{eqnarray}
with
\begin{eqnarray}
 c^h_{11} &=& C_F \left[
2\frac{\uuu}{\ttt-\mtsq}x_t\ln\frac{x_t}{1-x_t} +
\ln(1-x_t)(-2 x_t + 8 \ln x_t + 6) - 6 \ln^2 (1-x_t)
\right.
\nn
\\
&&
\left. - 4 {\rm Li}_2 \left( \frac{x_t}{x_t-1}\right)
-4 \ln^2 x_t + (2 x_t - 12)\ln x_t + \frac{29}{6}\pi^2 -
28\right],
\end{eqnarray}
where $x_t= \mtsq/\sss$. We have checked this result against
the existing NLO virtual corrections to s-channel single top
production~\cite{Harris:2002md} and found complete
agreement.
The remaining diagrams of Fig.~\ref{box} corresponds to
$H^1_{12}$ and $H^1_{21}$. They do not
contribute to the cross section at the NLO but do at the
NNLO, therefore are required for a complete NNLL
resummation. The results are
\begin{equation}
 H^1_{12}=H^1_{21}= \frac{1}{2}\gamma_{\rm cusp}^0
\ln\frac{\mu^2}{m^2_t}\ln\frac{\uuu(\uuu-\mtsq)}{
\ttt(\ttt-\mtsq)} + c^h_{12},
\end{equation}
with
\begin{eqnarray}
&&\frac{\hat{t}(\hat{t}-m^2_t)}{\hat{s}-M^2_W}c^h_{12}
\nn
\\
&=&
 -2 \left[\mtsq (\sss+2\ttt)+M^2_W \sss - \sss^2 - 2\sss
\ttt - 2\ttt^2 \right] C_0(0,0,\sss,M^2_W,0,0)
\nn
\\
&&
+\frac{1}{\sss+\ttt} \left[ \mtsq ( \ttt (3\ttt-2 M^2_W) +
\sss^2 + 4\sss\ttt)+ (\sss+\ttt)^2 (M^2_W - \sss - 2\ttt)
\right] C_0 (0,\uuu,\mtsq,M^2_W,0,0)
\nn
\\
&&
+ 4 \ttt (\mtsq - \ttt) C_0(0,0,\sss,0,0,M^2_W)
- \uuu (M^2_W + \uuu - \ttt) C_0(0,0,\uuu,0,0,0)
\nn
\\
&&
+ (\mtsq-\sss)(M^2_W + \uuu - \ttt)
C_0(0,\sss,\mtsq,0,0,M^2_W)
- \uuu ( M^2_W + \uuu - \ttt) C_0(0,\uuu,0,M^2_W,0,\mtsq)
\nn
\\
&&
+ (\sss+\ttt) (M^2_W + \uuu
-\ttt)C_0(\mtsq,0,\uuu,\mtsq,0,0)
+ 2\mtsq \ttt C_0(0,\ttt,0,M^2_W,0,\mtsq)
\nn
\\
&&
\frac{1}{\mtsq-\sss} \left[ m^6_t + m^4_t ( M^2_W -
3\sss-2\ttt) + m^2_t ( 3\sss
(\sss+2\ttt)-2M^2_W(\sss+\ttt)) + \sss^2 (M^2_W
-\sss-2\ttt)\right]
\nn
\\
&&
\times  C_0(\mtsq,\sss,0,\mtsq,0,M^2_W)
- 2 \ttt^2 (\mtsq - \ttt)
D_0(0,0,0,\mtsq,\ttt,\sss,0,0,0,M^2_W)
\nn
\\
&&
2\ttt (\mtsq - \ttt)^2 D_0
(0,\sss,0,\ttt,\mtsq,0,\mtsq,M^2_W,0,0)
\nn
\\
&&
+ (\sss+\ttt)\left[ \mtsq (M^2_W-\sss-2\ttt)+M^4_W-2 M^2_W
(\sss+\ttt) + \sss^2 + 2\sss\ttt+2\ttt^2\right]
\nn
\\
&&
\times D_0(\mtsq,0,0,0,\uuu,\sss,\mtsq,0,0,M^2_W)
\nn
\\
&&
- \Big[ m^4_t ( M^2_W -\sss-2\ttt) + \mtsq ( M^4_W - 3
M^2_W (\sss+\ttt) + 2\sss^2 + 5\sss\ttt + 4\ttt^2)
\nn
\\
&&
- M^4_W (\sss+\ttt) + 2M^2_W (\sss+\ttt)^2 - \sss^3 -
3\sss^2\ttt - 4\sss\ttt^2 - 2\ttt^3 \Big] D_0(
0,0,0,\mtsq,\sss,\uuu,M^2_W,0,0,0)
\nn
\\
&&
+ \frac{ 2\mtsq \ttt (\uuu-\sss)}{(\mtsq-\sss)(\sss+\ttt)}
(B_0 (\mtsq,0,M^2_W) + B_0(\mtsq,0,\mtsq))
+\frac{4 \sss\ttt}{\mtsq-\sss} B_0(\sss,0,M^2_W)
\nn
\\
&&
-\frac{2\ttt\uuu}{\sss+\ttt}( B_0(\uuu,0,0) +
B_0(\uuu,0,\mtsq)) - \frac{2}{\epsilon_{\rm IR}}
\frac{\hat{t}(\hat{t}-m^2_t)}{\hat{s}-M^2_W}\ln \frac{\uuu
(\uuu-\mtsq)}{\ttt ( \ttt-\mtsq)},
\end{eqnarray}
where the $B_0$, $C_0$ and $D_0$ are the conventional
Passarino-Veltman function~\cite{Passarino:1978jh},
evaluated at the point where the 't Hooft mass is set to
$m_t$. For example, the $B_0$ function reads
\begin{equation}
 B_0 (p^2_1,
m^2_1,m^2_2)=\frac{m^{2\epsilon}_t\Gamma(1-\epsilon)}{i\pi^{
2-\epsilon }}\int\, d^{4-2\epsilon} l
\,\frac{1}{(l^2-m^2_1+i\varepsilon)((l+p_1)^2-m^2_2 +
i\varepsilon)},
\end{equation}
and similarly for $C_0$ and $D_0$. This is just a simple
way to extract the $\mu$ dependece from $H^1_{12}$. The
analytical form of the singular Passarino-Veltman function
can be found, {\it
e.g.}, in the Ref.~\cite{Ellis:2007qk}. Note that
$H^1_{12}$ and $H^1_{21}$ are UV and
IR finite. Given the 1-loop matching coefficient above, one
can check that the RG evolution equation of the hard
function exactly has the form of Eq.~(\ref{hardrg}).

The calculation of the soft function can be divided into
the calculations of soft integral $I_S$ and the
corresponding color factor. We first discuss the soft
integral below. The soft integral corresponding to diagram
(a) of Fig.~\ref{softint} and its mirror image counterpart
can
be written as
\begin{equation}
 I_a =  2 g^2_s
\left(\frac{\mu^2 e^{\gamma_E}}{4\pi} \right)^\epsilon \int
\, \frac{d^n q}{(2\pi)^{n-1}} \delta (q^2) \theta(q_0)
\delta(k^+ - n_1 \cdot q) \frac{n_a \cdot n_b}{(n_a \cdot
q)(n_b \cdot q)},
\end{equation}
 where we work in $n=4-2\epsilon$ dimension, and the factor
of 2 comes from doubling the contribution of diagram (a) by
including its mirror image counterpart. This integral is
evaluated by Becher and Schwartz~\cite{Becher:2009th},
with the result:
\begin{equation}
 \bar{I}_a = \frac{\alpha_s}{4\pi} \left\{ \left[ 2 \ln^2
\frac{2 n_{ab}}{n_{1a}n_{1b}} - \frac{\pi^2}{3} \right]
\delta(k^+) + 16\left[ \frac{1}{k^+} \ln\left(
\frac{k^+}{\mu}
\sqrt{\frac{2 n_a\cdot n_b}{n^+_{a}n^+_{b}}}\right)
\right]^{[k^+,\mu]}_\star\right\},
\end{equation}
where $n^+_{a(b)}=n_{a(b)} \cdot n_1$ and
$[f]^{[a,b]}_\star$ is the star distribution defined in the
Ref.~\cite{Schwartz:2007ib}. Note that we have put a bar
on $I$ to denote that divergent terms have been subtracted
in $\overline{\text{MS}}$ scheme.

The soft integral corresponding to diagram (b) and (c) of
Fig.~\ref{softint} reads
\begin{eqnarray}
\label{ib}
 I_b &=& - g^2_s \left(\frac{\mu^2 e^{\gamma_E}}{4\pi}
\right)^\epsilon\int\,\frac{d^n q}{(2\pi)^{n-1}} \delta
(q^2) \theta(q_0)
\delta(k^+ - n_1 \cdot q) \frac{1}{(v\cdot q)^2},
\\
\label{ic}
I_c &=& 2 g^2_s \left(\frac{\mu^2 e^{\gamma_E}}{4\pi}
\right)^\epsilon\int\,\frac{d^n q}{(2\pi)^{n-1}} \delta
(q^2) \theta(q_0)
\delta(k^+ - n_1 \cdot q) \frac{v^+}{(n_1\cdot q)(v\cdot
q)},
\end{eqnarray}
where $v^+ = v\cdot n_1$. The simplest way to do the
integral of $I_b$ and $I_c$ is working in the lightlike
coordinates along the $n_1$ direction, in which any four
vector
can be written as
\begin{equation}
 p^\mu = \frac{1}{2}p^- n^\mu_1 + \frac{1}{2} p^+
\bar{n}^\mu_1 +
p^\mu_\perp.
\end{equation}
The results for the integrals in Eqs.~(\ref{ib}) and
(\ref{ic}) are
\begin{eqnarray}
 \bar{I}_b &=& \frac{\alpha_s}{4\pi} \left\{ 4\ln v^+
\delta(k^+) - 4 \left[
\frac{1}{k^+}\right]^{[k^+,\mu]}_\star\right\},
\\
\bar{I}_c &=& \frac{\alpha_s}{4\pi}  \left\{
\left[ -4\ln^2 v^+ - \frac{\pi^2}{6}\right] \delta(k^+) -
8 \left[ \frac{1}{k^+}
\ln\frac{k^+}{v^+\mu}\right]^{[k^+,\mu]}_\star\right\}.
\end{eqnarray}

The next two diagrams, (d) and (e), vanish, as was
explained in the Ref.~\cite{Becher:2009th}. Diagrams
(f) and (g) are the most complicated diagrams to be
evaluated. They read
\begin{eqnarray}
\label{intf}
 I_f &=& 2 g^2_s \left(\frac{\mu^2 e^{\gamma_E}}{4\pi}
\right)^\epsilon\int\,\frac{d^n q}{(2\pi)^{n-1}} \delta
(q^2) \theta(q_0)
\delta(k^+ - n_1 \cdot q) \frac{n_a \cdot v}{(n_a\cdot
q)(v\cdot
q)},
\\
I_g &=&- 2 g^2_s \left(\frac{\mu^2 e^{\gamma_E}}{4\pi}
\right)^\epsilon\int\,\frac{d^n q}{(2\pi)^{n-1}} \delta
(q^2) \theta(q_0)
\delta(k^+ - n_1 \cdot q) \frac{n_b \cdot v}{(n_b\cdot
q)(v\cdot
q)}.
\end{eqnarray}
We calculate diagram (f) first. As before, we work in a
lightlike coordinates along the $n_1$ direction. The delta
function in Eq.~(\ref{intf}) can be used to integrate out
$dq^+$ and $dq_\perp$. After some simplification we arrive
at
\begin{equation}
\label{intf1}
 I_f = \frac{\alpha_s}{4\pi}
\frac{(\mu v^+)^{2\epsilon}}{(k^+)^{1+2\epsilon}}
(1+\rho_a)\frac{4
e^{\epsilon\gamma_E}}{\sqrt{\pi}\Gamma(1/2-\epsilon)}
\int^\infty_0 \, dx \int^\pi_0\, \frac{d
\theta}{\sin^{2\epsilon}\theta}
\frac{x^{-\epsilon}}{(1+x)(x+\rho_a-2\sqrt{x\rho_a}
\cos\theta) } ,
\end{equation}
where we have defined $\rho_a=\frac{(v^+)^2 n^-_a}{n^+_a}$,
$x=\frac{q^- (v^+)^2}{k^+}$, and $\theta$ is the angle
between $\vec{q}_\perp$ and $\vec{n}_{a\perp}$. Note that
we have $(1+\rho_a)=2 n_a\cdot v \frac{
v^+}{n^+_a}$. A somewhat similar integral appear in the
evaulation of hadronic thrust
distribution~\cite{Kelley:2010qs}.
The integral in Eq.~(\ref{intf1}) is straightfoward to do,
with a result
\begin{equation}
 I_f = \frac{\alpha_s}{4\pi}
\frac{(\mu v^+)^{2\epsilon}}{(k^+)^{1+2\epsilon}}\left[
-\frac{4}{\epsilon} + 8\ln(1+\rho_a) + \epsilon \left(
8{\rm Li}_2\left(\frac{\rho_a}{1+\rho_a}\right) -
4\ln^2(1+\rho_a)-\frac{\pi^2}{3}\right)\right].
\end{equation}
Using the expansion of $(k^+)^{-1-2\epsilon}$ in terms of
star distribution~\cite{Schwartz:2007ib}:
\begin{equation}
 \frac{1}{k^+}\left(
\frac{\mu}{k^+}\right)^{2\epsilon} =
-\frac{1}{2\epsilon}\delta(k^+) + \left[
\frac{1}{k^+}\right]^{[k^+,\mu]}_\star - 2\epsilon \left[
\frac{1}{k^+}\ln\frac{k^+}{\mu}
\right]^{[k^+,\mu]}_\star+\mathcal{O}(\epsilon^2),
\end{equation}
we obtain
\begin{eqnarray}
 \bar{I}_f &=& \frac{\alpha_s}{4\pi} \left\{ \left[ -4 {\rm
Li}_2 \left(\frac{\rho_a}{1+\rho_a}\right) +
2\ln^2(1+\rho_a) -8\ln v^+ \ln (1+\rho_a) +4 \ln^2v^+
+\frac{\pi^2}{6} \right] \delta(k^+)
\right.
\nn
\\
&&
\left.+ 8\left[ \frac{1}{k^+} \ln
\left(\frac{k^+}{\mu}\frac{1+\rho_a}{v^+}
\right)\right]^{[k^+,\mu]}_\star\right\}.
\end{eqnarray}
Diagram (g) can be obtained from diagram (f) by the simple
replacement $\bar{I}_g = -\bar{I}_f (\rho_a \to \rho_b)$.
Given the integral above, the soft function can be derived
by combining them with the corresponding color factor:
\begin{eqnarray}
\frac{\alpha_s}{4\pi}{S}^1_{11} &=& C^2_A C_F
(\bar{I}_a+\bar{I}_b+\bar{I}_c),
\nn
\\
\frac{\alpha_s}{4\pi}{S}^1_{12} &=& \frac{1}{2}C_A
C_F (\bar{I}_{f}+\bar{I}_g),
\nn
\\
{S}^1_{21} &=&
{S}^1_{12},
\nn
\\
\frac{\alpha_s}{4\pi}{S}^1_{22} &=& -\frac{1}{4} C_F
(\bar{I}_a+\bar{I}_c) + \frac{1}{2}C_A C^2_F \bar{I}_b
+\frac{1}{4}C_F(C^2_A-2) \bar{I}_f - \frac{1}{2}C_F
\bar{I}_g.
\end{eqnarray}
We note that the complete set of color factor for all $2\to
2$ processes has been worked out in the
Ref.~\cite{Kelley:2010qs}. To check that the soft
function we obtain indeed obeys the RG evolution
Eq.~(\ref{sevol}), it's convinient to make a Laplace
transformation to the soft function. Explicitly, we have
\begin{eqnarray}
 \widetilde{s}^1_{11}(L,\mu)&=&C^2_A C_F \left( 4 L^2 +
8
\ln\frac{2 n_a \cdot n_b v^+}{n^+_a n^+_b} L - 4 L \right)
+ c^s_{11},
\nn
\\
\widetilde{s}^1_{12}(L,\mu) &=& 4 C_A C_F
\ln\frac{1+\rho_a}{1+\rho_b} L + c^s_{12},
\nn
\\
\widetilde{s}^1_{21}(L,\mu)&=&\widetilde{s}^1_{12}(L
, \mu),
\nn
\\
\widetilde{s}^1_{22}(L,\mu)&=& C_F (C^2_A - 1) L^2 - 2
C_A
C^2_F L - 2 C_F \ln\frac{2 n_a \cdot n_b v^+}{n^+_a n^+_b} L
\nn
\\
&&
+ 2 C_F (C^2_A - 2) \ln\frac{1+\rho_a}{v^+} L
 + 4 C_F
\ln\frac{1+\rho_b}{v^+} L + c^s_{22},
\end{eqnarray}
where $L=\ln\frac{\kappa}{\mu}$. and
\begin{eqnarray}
 c^s_{11} &=& C^2_A C_F \left( 2\ln^2\frac{2
n_{ab}}{n_{1a}n_{1b}} - 4\ln^2 v^++4 \ln v^+ +
\frac{\pi^2}{6} \right),
\nn
\\
c^s_{12} &=& C_A C_F \left( -2 {\rm Li}_2 \left(
\frac{\rho_a}{1+\rho_a}\right) + 2{\rm Li}_2
\left(\frac{\rho_b}{1+\rho_b}\right) - 4\ln v^+
\ln(1+\rho_a) + 4\ln v^+\ln(1+\rho_b)
\right.
\nn
\\
&&
 + \ln^2(1+\rho_a) -
\ln^2(1+\rho_b)\Big),
\nn
\\
c^s_{22} &=& C_F C^2_A \left( -{\rm Li}_2 \left(
\frac{\rho_a}{1+\rho_a}\right) - 2\ln v^+ \ln (1+\rho_a) +
\ln^2 v^+ + \frac{1}{2} \ln^2 (1+ \rho_a) +
\frac{5}{24}\pi^2 \right)
\nn
\\
&&
+C_F \left( -\frac{1}{2} \ln^2\frac{2 n_{ab}}{n_{1a} n_{1b}}
+ 2{\rm Li}_2 \left(
\frac{\rho_a}{1+\rho_a}\right) - 2{\rm Li}_2 \left(
\frac{\rho_b}{1+\rho_b}\right) + 4\ln v^+ \ln(1+\rho_a)
\right.
\nn
\\
&&
\left.
 -
4\ln v^+ \ln(1+\rho_b) + \ln^2 v^+  - \ln^2 (1+\rho_a) +
\ln^2 (1+ \rho_b) - \frac{\pi^2}{24} \right) + 2C_A C_F^2
\ln v^+.
\end{eqnarray}
Using the expressions
above, we confirm that the RG equation of the
soft function agrees with Eq.~(\ref{sevol}). This shows
that our resummed cross section is RG invariant in the
threshold limit, which can be considered as a non-trivial
check of our result.

Finally, the Laplace transformed jet function is given
by~\cite{Becher:2006qw}:
\begin{equation}
 \widetilde{j}(L,\mu)=1+\frac{\alpha_s(\mu)}{4\pi}\left[
\frac{1}{2}C_F \gamma^0_{\rm cusp} L^2 + \gamma^0_j L +
c^j_1\right],
\end{equation}
with $c^j_1 = C_F\times\left( 7- \frac{2}{3}\pi^2 \right)$.

\bibliography{stop}{}
\end{document}